\documentclass[11pt]{article}
\usepackage[utf8]{inputenc}
\usepackage[T1]{fontenc}
\usepackage[DIV12]{typearea}
\usepackage{amsmath,amsfonts,amssymb,amscd}
\usepackage[all]{xy}
\usepackage{graphicx}
\usepackage{cite}
\usepackage{mathrsfs}
\usepackage{bbm}
\usepackage{bbold}
\usepackage{braket}
\usepackage[dvipsnames]{xcolor}
\usepackage[normalem]{ulem}
\usepackage{tensor}
\usepackage{mathtools}
\usepackage[bookmarks=false]{hyperref}
\usepackage{srcltx}

\newcommand{\be}{\begin{equation}}
\newcommand{\ee}{\end{equation}}
\newcommand{\bea}{\begin{eqnarray}}
\newcommand{\eea}{\end{eqnarray}}

\DeclareMathOperator{\WHSIC}{WH-SIC}
\DeclareMathOperator{\WHMUB}{WH-MUB}

\newcommand{\E}{\mathcal{E}}

\newcommand{\B}{\mathcal{B}}

\newcommand{\ketbra}{\rangle \langle}
\def\ket{\rangle}
\def\bra{\langle}

\def\b{\beta}

\def\B4{{\cal B}_{442}}

\def\eB4{\bra {\cal B}_{442} \ket}

\usepackage{tikz}
\usetikzlibrary{quantikz2}

\def\bea{\begin{eqnarray}}
\def\eea{\end{eqnarray}}
\def\be{\begin{equantion}}
\def\ee{\end{equation}}

\def\be{\begin{equation}}
\def\ee{\end{equation}}

\def\ket{\rangle}
\def\bra{\langle}

\def\E8{{\rm E}_8}
\def\BW16{{\rm BW}_{16}}

\definecolor{dg}{rgb}{0,.5,0}

\usepackage{soul}

\ifdefined\SHOWCHANGES
  \newcommand{\rev}[1]{{\color{red}#1}}
  \newenvironment{revision}{\color{red}}{}
\else
  \newcommand{\rev}[1]{#1}
  \newenvironment{revision}{}{}
\fi

\def\mytitle{Extremal Magic States from Symmetric Lattices}

\title{\mytitle}

\begin{document}

\begin{titlepage}
\setlength{\topmargin}{0.0 true in}
\thispagestyle{empty}

\vspace{-10mm}


\vspace{10mm}

\begin{center}
{\LARGE\textbf{\mytitle}}

\renewcommand*{\thefootnote}{\fnsymbol{footnote}}

\vspace{1.1cm}
\large
Misaki Ohta${}^{1}$
~and~
Kazuki Sakurai${}^{2}$
\\[5mm]

\normalsize

\textit{
${}^1$
Institute of Theoretical Physics, University of Wroc\l{}aw, \\ plac Maksa Borna 9, PL-50204 Wroc\l{}aw, Poland
} \\[2mm]
\textit{
${}^2$Institute of Theoretical Physics, Faculty of Physics,\\
University of Warsaw, Pasteura 5, 02-093, Warsaw, Poland
} \\[2mm]

\end{center}
\vspace*{5mm}

\begin{abstract}\noindent

Magic, a key quantum resource beyond entanglement, remains poorly understood in terms of its structure and classification. 
In this paper, we demonstrate a striking connection between high-dimensional symmetric lattices and quantum magic states.
By mapping vectors from the $E_8$, $BW_{16}$, and $E_6$ lattices into Hilbert space, we construct and classify stabiliser and maximal magic states for two-qubit, three-qubit and one-qutrit systems. 
\rev{In particular, this geometric approach gives closed-form lattice representatives of maximal magic states and motivates a conjecture for the number of three-qubit maximal states.}
In the three-qubit case, we further classify the extremal magic states according to their entanglement structure.
\rev{We also show that the 45 one-qutrit maximal magic states selected by the second shell of $E_6$ split into two Clifford orbits.}
Our findings suggest that deep algebraic and geometric symmetries underlie the structure of extremal magic states.

\end{abstract}

\end{titlepage}

\section{Introduction}

One of the most striking insights in quantum information theory is the Gottesman–Knill theorem \cite{gottesman1998,Gottesman:1999tea,Aaronson:2004xuh}, which states that any quantum computation restricted to Clifford unitaries and stabiliser inputs can be simulated efficiently on a classical computer. This implies that states lying outside the stabiliser polytope—so-called magic states \cite{Bravyi:2004isx}—are essential for achieving quantum computational advantage. 
While entanglement has traditionally been viewed as a central quantum resource, magic embodies a complementary form of nonclassicality that remains largely unexplored.

Recent efforts to quantify non-stabiliser resources have led to a variety of proposed measures \cite{Mari,Veitch_2012,Veitch_2014, Bravyi2019, SRE, Warmuz:2024cft}, among which the stabiliser Rényi entropies (SREs) of order $\alpha$ \cite{SRE}, denoted $M_\alpha(|\psi\rangle)$, have emerged as particularly useful. The SREs are well-defined for general $n$-qudit systems \cite{Wang2023}, experimentally measurable \cite{Oliviero:2022bqm,Haug:2023ffp} and computationally tractable. Notably, it has been shown that SREs with $\alpha \geq 2$ behave as resource monotones under the stabiliser protocol—operations composed of Clifford unitaries and projective measurements—when restricted to pure states \cite{Leone:2024lfr}. This implies that $M_{\alpha \geq 2}(|\psi\rangle)$ quantifies a genuine resource for quantum computation, reinforcing its relevance as a measure of magic.

While SREs provide a useful quantification of magic, the maximal value of $M_\alpha(|\psi\rangle)$ and the states that achieve it are not yet fully understood.
Recently, upper bounds for $M_{\alpha \geq 2}(|\psi\rangle)$ have been established for general $D$-dimensional Hilbert spaces \cite{Wang2023, Cuffaro:2024wet}, with equality if and only if $|\psi\rangle$ belongs to a class of Weyl–Heisenberg Covariant Symmetric Informationally Complete (WH-SIC) states. However, the existence of WH-SIC states in all finite dimensions remains an open problem \cite{axioms6030021}. Although WH-SIC states are believed to exist in all single-qudit systems, their existence has been ruled out for composite $n$-qubit systems (i.e., ($\mathbb{C}^2)^{\otimes n}$) except when $n = 3$ \cite{axioms6030021}. More recently, a tighter bound for the two-qubit case ($n = 2$) was identified \cite{Liu:2025frx}, saturated only by a special set of states known as WH-MUBs, which generate Mutually Unbiased Bases as orbits of the Weyl–Heisenberg group. The authors conjecture that WH-MUBs saturate the optimal bound for general $n$-qubit systems, except when $n = 1$ or $n = 3$.
The fact that the sets of maximal magic states form WH-SICs or WH-MUBs in their Hilbert spaces may suggest that these states generally exhibit particular geometric patterns in certain high-dimensional spaces.

Despite these developments, closed-form expressions and complete enumerations of maximal magic states are known only for the single-qubit and two-qubit cases, while three-qubit and higher-dimensional systems continue to pose significant open questions.

In this work, we demonstrate a remarkable geometric connection between extremal magic states and symmetric high-dimensional lattices: the $E_8$, $BW_{16}$, and $E_6$ lattices. 
By mapping lattice vectors into Hilbert space, we systematically construct and classify stabiliser and maximal magic states in the two-qubit, three-qubit and one-qutrit systems. 
\rev{In particular, we obtain closed-form lattice representatives of maximal magic states in the three-qubit and one-qutrit cases \cite{ohta_2025_15651358}.}

For the three-qubit system, we propose a conjecture that the total number of maximal magic states is 15,360. 
We also provide a detailed classification of those states based on their entanglement structure.
\rev{In the one-qutrit system, we identify 45 maximal magic states selected by the second shell of $E_6$ and show that this finite lattice-selected set partitions into two distinct Clifford orbits. This statement concerns the $E_6$ shell rather than the full, continuous family of one-qutrit WH-SIC fiducials \cite{TabiaAppleby2013}.}

The remainder of the paper is organised as follows.
In the next section, we introduce a mapping from lattice vectors to vectors in Hilbert space, which forms the foundation for our main results.
Based on this construction, we summarise and enumerate the main findings of the work.
Sections 3 and 4 provide preliminary background on extremal magic states and lattice theory, respectively.
Sections 5, 6 and 7 are devoted to detailed discussions of the two-qubit, three-qubit and one-qutrit systems.
Finally, Section 8 presents our conclusions.

\section{Summary of main findings}

\subsection{The map from lattice points to quantum states}
\label{sec:map}

The main findings of this work are based on simple functions that send a lattice vector in ${\mathbb C}^D$ or ${\mathbb R}^{2D}$ to a pure quantum state in the $D$-dimensional Hilbert space, ${\cal H} = {\mathbb C}^D$.
When a lattice vector $\vec c = (c_1,\ldots,c_D)$ resides in ${\mathbb C}^D$, we obtain the corresponding quantum state through a map, $g$, defined by 
\be
\vec c \, \xlongrightarrow{g} \, | \psi \ket = 
\frac{1}{\cal N} \sum_{k=1}^D c_k | k \ket,
~~~~
{{\cal N} = \sqrt{\sum_{k=1}^D |c_k|^2}},
\ee
where $\{ | k \ket \}_{k=1}^D$ is the orthonormal computational basis in ${\cal H} = {\mathbb C}^D$.
Note that this map is not invertible: different vectors $\vec{c}$ and $\vec{c}^{\,\prime}$ can be mapped to the same quantum state if they differ by a complex scaling factor, i.e., $\vec{c}^{\,\prime} = u \vec{c}$ for some $u \in \mathbb{C}$.

If instead the lattice vector $\vec{x} = (x_1, \ldots, x_{2D})$ lies in $\mathbb{R}^{2D}$, we first construct a complex vector $\vec{c} \in \mathbb{C}^D$ using the bijective map $\varphi: \mathbb{R}^{2D} \rightarrow \mathbb{C}^D$, defined by:
\be
\vec{x} = 
(x_1, \ldots, x_D, x_{D+1},\ldots, x_{2D})
\xlongrightarrow{\varphi} 
\vec c =
(c_1, \ldots, c_D),
~~~~
c_k = x_k + i x_{D+k}\,.
\label{map_phi}
\ee
The corresponding quantum state is then given by composing $g$ with $\varphi$:
$| \psi \rangle = g\big( \varphi(\vec{x}) \big)$.

A symmetric lattice defines an infinite set of points, $\{ \vec{c}_k \}$ in $\mathbb{C}^{D}$ or $\{ \vec{x}_k \}$ in $\mathbb{R}^{2D}$, generated by a finite number of basis vectors.
These vectors are classified into ordered sets according to their norms, given by
$\langle \vec{c}, \vec{c} \rangle = \sum_{k=1}^{D} |c_k|^2$ or $\langle \vec{x}, \vec{x} \rangle = \sum_{k=1}^{2D} x_k^2$. 
Specifically, the sets of lattice vectors with the smallest, second-smallest, and third-smallest norms are referred to as the sets of first-shortest, second-shortest, and third-shortest vectors, respectively, and so on.\footnote{Symmetric lattices possess the unique null vector at the origin.  We do not consider this null vector in our consideration.}

By applying the map $g$ or $g \circ \varphi$, these sets of shortest vectors are translated into corresponding sets of pure quantum states:
$\{ |\psi_k\rangle = g (\vec{c}_k) \mid \bra \vec c, \vec c \ket = \ell_n \} $ ($\{ |\psi_k\rangle = g \circ \varphi (\vec{x}_k) \mid \bra \vec x, \vec x \ket = \ell_n \} $), where $\ell_n$ denotes the known squared norm of the $n$-th shortest vectors in the lattice. 
These structured sets of states form the basis for our investigation into the minimal and maximal magic states in quantum systems.

\subsection{Summary of main findings}

We summarise the main findings of the paper. 

\subsection*{Results for $E_8$ lattice and two-qubit $({\mathbb C}^2)^{\otimes 2}$}
\begin{description}

\item[1.1] 
All 240 first-shortest vectors of the $E_8$ lattice are mapped to the complete set of 60 stabiliser states in the two-qubit system.

\item[1.2]
Out of the 2,160 second-shortest vectors of the $E_8$ lattice, 1,920 are mapped to the complete set of 480 previously known maximal magic states in the two-qubit system, while the remaining 240 vectors are mapped to the 60 stabiliser states.

\end{description}

\subsection*{Results for $BW_{16}$ lattice and three-qubit $({\mathbb C}^2)^{\otimes 3}$}
\begin{description}

\item[2.1] 
All 4,320 first-shortest vectors of the $BW_{16}$ lattice are mapped to all known 1080 stabiliser states of the three-qubit system.

\item[2.2]
All 61,440 second-shortest vectors of the $BW_{16}$ lattice are mapped to a set of 15,360 previously unreported maximal magic states in the three-qubit system.
Based on the symmetry properties of the lattice and the structure of the mapping, we conjecture that this set exhausts all maximal magic states in the three-qubit system.

\item[2.3]
The stabiliser and maximal magic states of the three-qubit system can be classified based on their entanglement properties. A summary of this entanglement classification is provided in Fig.\ \ref{magic-entangle}.

\end{description}

\subsection*{Results for $E_6$ lattice and single-qutrit ${\mathbb C}^3$}
\begin{description}

\item[3.1] 
All 72 first-shortest vectors of the $E_6$ lattice are mapped to all 12 stabiliser states of the single-qutrit system.

\item[3.2]
\rev{All 270 second-shortest vectors of the $E_6$ lattice are mapped to a finite lattice-selected set of 45 maximal magic states in the single-qutrit system. This does not exhaust the continuous family of one-qutrit WH-SIC fiducials \cite{TabiaAppleby2013}.}

\item[3.3]
\rev{The 45 $E_6$-derived maximal magic states are partitioned into two Clifford orbits with 9 and 36 distinct quantum states \cite{Appleby2005,ohta_2025_15651358}.}  

\end{description}

\begin{revision}
The common mechanism is group-theoretic.  In the complex coordinates used here, direct generator checks show that the relevant Clifford groups preserve the three lattices, so each norm shell is a union of Clifford orbits \cite{ohta_2025_15651358}.  For $BW_{16}$, this is an instance of the classical relation between Barnes--Wall automorphisms and Clifford groups \cite{NebeRainsSloane2001,kliuchnikov2024stabilizeroperatorsbarneswalllattices}.  For a representative in the appropriate second shell, the Weyl--Heisenberg overlaps satisfy the WH-MUB or WH-SIC equality condition; Clifford symmetry then propagates this property over its orbit \cite{axioms6030021,SzymusiakSlomczynski2016}.  Concretely, the relevant shell components have 480 WH-MUB rays for $E_8$, 15,360 WH-SIC fiducial rays for $BW_{16}$, and two WH-SIC fiducial orbits of 36 and 9 rays for $E_6$.  These orbit sizes agree with the corresponding shell counts after scalar unit phases are removed, which explains why these three lattices isolate the extremal families rather than merely producing them by numerical coincidence \cite{conway2013sphere,ohta_2025_15651358}.
\end{revision}

The coordinates of the first- and second-shortest vectors of the $E_8$, $BW_{16}$ and $E_{6}$ lattices, along with the stabiliser and maximal magic states in the two-qubit, three-qubit and one-qutrit systems—classified by their entanglement properties and Clifford orbits—are provided in \cite{ohta_2025_15651358}. 
This resource also includes scripts for computing the stabiliser R{\' e}nyi entropy and various entanglement measures.

\section{Preliminaries: Extremal magic states}
\label{sec:magic}

Magic is the concept that characterises the non-stabilizerness of the quantum state. 
The stabiliser R{\'e}nyi entropis (SREs) of order $\alpha$, $M_\alpha( | \psi \ket )$, defined by
\be
M_\alpha( | \psi \ket ) \,\equiv\,
- \frac{1}{\alpha - 1}
\log \Xi_\alpha( | \psi \ket )
\label{defM0}
\ee
is particularly useful to quantify the magic. 
Unless otherwise specified, all logarithms in this paper are taken to base 2.
For the general $n$-qudit system $({\mathbb C}^d)^{\otimes n}$, 
\be
\Xi_\alpha(| \psi \ket) \,\equiv\, \frac{1}{d^n} \sum_{ {\cal O} \in {\cal W}(n,d)} \left| \bra \psi | {\cal O} | \psi \ket \right|^{2 \alpha} ,
\label{Xi}
\ee
where ${\cal W}(n,d)$ is the set of elements of the Weyl-Heisenberg (WH) group for $({\mathbb C}^d)^{\otimes n}$, $\overline {\cal W}(n,d)$,  with the trivial phase. 
Further information regarding the WH group and the SRE is provided in Appendix \ref{SRE}.
It has been demonstrated that the SREs of order $\alpha$ exhibit particularly favourable properties when $\alpha \geq 2$.
In the following, we therefore set $\alpha = 2$ for concreteness and simplicity.

The Clifford group, ${\cal C}(n,d)$, for $({\mathbb C}^d)^{\otimes n}$ is defined as the normaliser of the WH group $\overline {\cal W}(n,d)$ in the unitary group $U(d^n)$
\be
{\cal C}(n,d) \equiv \{ 
U \in U(d^n) \,|\, U^\dagger \overline {\cal O} U \in
\overline {\cal W}(n,d),~
^\forall \overline {\cal O} \in \overline {\cal W}(n,d)
\} \,.
\ee
As the Clifford unitaries send an element of ${\cal W}(n,d)$ to another element of ${\cal W}(n,d)$, it leaves $\Xi_\alpha(| \psi \ket)$ 
(and therefore $M_\alpha( | \psi \ket )$) invariant. 
Since the Clifford group contains multitudes of phase-equivalent elements, it is often convenient to consider the quotient group ${\cal C}(n,d)/U(1)$. 
For the $n$-qubit system, $({\mathbb C}^2)^{\otimes n}$, the order of this group is 
$\left| {\cal C}(n,2)/U(1) \right| = 2^{n(n+2)} \Pi_{j=1}^n (2^j - 1)$ \cite{Mastel:2023uhk}, while for the single-qudit, ${\mathbb C}^d$, $\left| {\cal C}(1,d)/U(1) \right| = d^3 (d^2 -1)$ \cite{Morvan_2021}.

In the $n$-qubit system $({\mathbb C}^2)^{\otimes n}$, ${\cal W}(n,2)$ reduces to the set of Pauli strings
\be
{\cal P}_n \,\equiv\, \{ P_1 \otimes \cdots \otimes P_n \,|\, P_i \in \{ 1, \sigma_x, \sigma_y, \sigma_z \}  \} 
\ee
and $\Xi_2(| \psi \ket)$ is given by
\be
\Xi_2(| \psi \ket) \,\equiv\, \frac{1}{2^n} \sum_{P \in {\cal P}_n} \left| \bra \psi | P | \psi \ket \right|^{4} \,.
\label{Xi2_nbits}
\ee

One way to make sense of the SRE is as follows \cite{Wang2023,Cuffaro:2024wet}.
For a given Pauli string $P \in {\cal P}_n$, define
\be
Q_P(| \psi \ket) = 
\frac{1}{2^n} \left| \bra \psi | P | \psi \ket \right|^{2} 
\ee
One can show, for any normalised $n$-qubit state $| \psi \ket$, $Q_P(| \psi \ket) \geq 0$ and  
$\sum_{P \in {\cal P}_n} Q_P(| \psi \ket) = 1$.
Namely, $Q_P(| \psi \ket)$ can be interpreted as a probability distribution. 
For this distribution, the $\alpha$-R{\' e}nyi entropy \cite{Rnyi1961OnMO} is defined as 
\be
M_\alpha( | \psi \ket )
\equiv \frac{1}{1 - \alpha} \log 
\sum_{P \in {\cal P}_n} \left[ Q_P(| \psi \ket) \right]^\alpha - \log 2^n\,,
\label{defM2}
\ee
which agrees with the SREs of order $\alpha$ defined in Eq.\ \eqref{defM0}.

The stabiliser states are defined as the states that are reachable from the computational basis $| 0 \ket^{\otimes n}$ by applying only the Clifford unitaries. 
In the $n$-qubit system $({\mathbb C}^2)^{\otimes n}$, the Clifford unitaries are constructed only using the $H$, $S$ and CNOT gates.
It has been shown that any quantum computation involving only stabiliser states has no quantum advantage \cite{gottesman1998,Gottesman:1999tea,Aaronson:2004xuh}.  
The stabiliser states are also defined by demanding that the corresponding stabiliser group, ${\cal S}_i$, stabilises them, such that
\be
S_i | \Psi_i \ket = | \Psi_i \ket,~~~^\forall S_i \in {\cal S}_i \,.
\label{stabiliser_def}
\ee
Here, the stabiliser groups are the maximal abelian subgroups of $\overline {\cal W}(n,d)$ that do not contain any non-trivial phase element other than 1. 
Since $|{\cal S}_i| = d^n \equiv D$, the above condition uniquely determines a quantum state in ${\cal H} = ({\mathbb C}^d)^{\otimes n}$ for a given ${\cal S}_i$.
For the $n$-qubit system, $({\mathbb C}^2)^{\otimes n}$, it has been known that the number of such groups is given by \cite{Aaronson:2004xuh}
\be
N_s = 2^n \prod_{k=0}^{n-1} \left( 2^{n-k} + 1\right)\,.
\label{Ns}
\ee


By construction, see Appendix \ref{SRE} for details, $M_2( | \psi \ket )$ takes the minimal value zero if and only if $| \psi \ket$ is a stabiliser state;
$| \psi \ket \in \{ | \Psi_i \ket \}$.

Our understanding of maximal magic states remains relatively limited. 
Recently, an upper bound on $M_\alpha(| \psi \rangle) (\alpha \geq 2)$ (and the corresponding lower bound on $\Xi_\alpha(| \psi \rangle)$) have been established. 
For general $D$-dimensional quantum systems, these bounds are expressed as:
\be
\Xi_{\alpha}( | \psi \ket )
\geq
\frac{1 + (D-1)(D + \Delta)^{1-\alpha}}{D}\,,
 ~~~~~~
M_{\alpha}( | \psi \ket ) \leq 
\frac{1}{1- \alpha} \log \frac{1 + (D-1)(D + \Delta)^{1-\alpha}}{D}
\label{bound}
\ee
with $\Delta = 1$.
Moreover, it has been shown that the bound is saturated if and only if the state belongs to a set of Weyl-Heisenberg covariant Symmetric Informationally Complete (WH-SIC) states.


A Symmetric Informationally Complete Positive Operator-Valued Measure (SIC-POVM) \cite{sic}, or SIC for short, is a set of $D^2$ rank-1 operators $\{ \Pi_i \equiv | \psi_i \ketbra \psi_i|  \}_{i=1}^{D^2}$ satisfying
\be
{\rm Tr} \left( \Pi_i \Pi_j \right) 
=
| \bra \psi_i | \psi_j \ket |^2 = 
\left\{
\begin{array}{cc}
1 & ~i=j \\
\frac{1}{D+1} & ~i \neq j
\end{array}
\right.
\label{sic}
\ee
By definition, a symmetric informationally complete (SIC) set spans the space of linear operators in ${\cal H} = \mathbb{C}^D$. When $\tfrac{1}{D} \Pi_i$ represents a POVM element in a measurement applied to a quantum state $\rho$, the associated probability is given by $p_i = \mathrm{Tr}[\Pi_i \rho]$.
Eq.\ \eqref{sic} indicates that the full set of probabilities $\{ p_i \}_{i=1}^{D^2}$ uniquely determines the state $\rho$, thereby justifying the term “informationally complete.”


A SIC is said to be covariant with respect to a group $G$ if 
\be
| \bra \psi_i | g | \psi_i \ket |^2 = 
\frac{1}{D+1},~~~~  \forall g \in G,~ g \neq{1} \,.
\label{wh-sic}
\ee
A WH-SIC is a set of states that satisfy both conditions \eqref{sic} and \eqref{wh-sic} with $G = \overline {\cal W}(n,d)$.
For general $D$-dimensional quantum systems, it is straightforward to see that if $| \psi \ket$ belongs to a WH-SIC, all $| \bra \psi | {\cal O} | \psi \ket |^{2 \alpha}$ terms in the summation of Eq.\ \eqref{Xi} produce $(D+1)^{-\alpha}$ except for the ${\cal O} = 1$ case.
As ${\cal W}(1,D)$ contains $D^2 - 1$ non-trivial elements (see Appendix \ref{SRE}), one can see that the bound \eqref{bound} with $\Delta = 1$ is saturated for a WH-SIC state.
For $\alpha = 2$, the bounds are written as
\be
\Xi_{2}( | \psi \ket )
\,\geq\,
\frac{2}{D+1}\,,
\,~~~~~~
M_{2}( | \psi \ket ) \,\leq\, 
\log \frac{D+1}{2}\,.
~~~\cdots~~ | \psi \ket \in {\WHSIC}
\label{bound_WHSIC}
\ee

WH-SICs do not always exist in general composite systems. 
For instance, for $n$-qubit systems, $({\mathbb C}^2)^{\otimes n}$, WH-SICs exist only for $n=1$ and $3$.
This means that for general composite $n$-qubit systems with $n \neq 3$, the upper bound \eqref{bound} is not saturated.

Recently, a tight upper bound for the 2-qubit system has been found, which is represented by the same expression as \eqref{bound} with $\Delta = 0$, regarding $D$ as the dimension of the total Hilbert space, i.e.\ $D = 2^2 = 4$ for 2-qubit.
The authors found that the bound is saturated if and only if $| \psi \ket$ is a fiducial state of WH-MUBs.
Two orthonormal bases $|\xi_i \ket$ and $|\eta_i \ket$ in ${\cal H} = {\mathbb C}^D$ are said to be mutually unbiased bases (MUBs) if $\bra \xi_i| \eta_j \ket = \frac{1}{D}$ for all $i,j = 1,\ldots,D$ \cite{mub} .
A state $| \psi \ket$ is said to be a WH-MUB fiducial state if the orbit of ${\cal O} | \psi \ket$ (${\cal O} \in {\cal W}(n, d)$) generates $D^2 = (d^{n})^2$ independent states that are partitioned into $D$ MUBs.
In this case, $D^2$ terms in the summation in \eqref{Xi} split into single 1, $(D-1)$ zeros 
and $(D-1)$ terms of $D^{-\alpha}$.
The bounds \eqref{bound} with $\Delta =0$ are, therefore, saturated for a WH-MUB fiducial state.
For $\alpha = 2$, they are expressed as
\be
\Xi_{2}( | \psi \ket )
\,\geq\,
\frac{2D-1}{D^2}\,,
\,~~~~~~
M_{2}( | \psi \ket ) \,\leq\, 
\log \frac{D^2}{2D-1}\,.
~~~\cdots~~ | \psi \ket \in {\WHMUB}
\label{bound_WHMUB}
\ee

\renewcommand{\arraystretch}{1.2} 
\begin{table}[t!]
\begin{center}
\resizebox{\textwidth}{!}{%
\begin{tabular}{| l | c | c | c | c |}
\hline
~System  & $\#$ stabiliser states & $\#$ lattice max-magic states & max magic states &  ($\Xi_2^{\rm min}$, $M_2^{\rm max}$)  \\ 
\hline
\hline
~1-qubit ${\mathbb C^2}$ & 6 & 8 & WH-SIC ($\Delta=1$) & ($\frac{2}{3},\, 0.585$) \\  
\hline
~2-qubit $({\mathbb C^2})^{\otimes 2}$ & 60 & 480 &  WH-MUB ($\Delta=0$) & ($\frac{7}{16},\, 1.193$)  \\
\hline
~3-qubit $({\mathbb C^2})^{\otimes 3}$ & 1080 & {\bf 15360} &  WH-SIC ($\Delta=1$) & ($\frac{2}{9}$, $2.170$)  \\
\hline
~1-qutrit $({\mathbb C^3})$ & 12 & {\bf 45} &  WH-SIC ($\Delta=1$) & ($\frac{1}{2}$, $1$)    \\
\hline
\end{tabular}%
}
\caption{\label{tab:magic_bounds} \small
\rev{Summary of the stabiliser states and the maximal magic states constructed in the indicated quantum systems \cite{ohta_2025_15651358}.}
The total count of the maximal magic states for the 2-qubit is found in \cite{Liu:2025frx}.
\rev{For the 3-qubit and 1-qutrit systems, the displayed numbers count the states explicitly constructed from the $BW_{16}$ and $E_6$ lattices, respectively; they should not be read as a classification of all maximisers in the ambient Hilbert space \cite{TabiaAppleby2013,ohta_2025_15651358}.}
}
\end{center}
\end{table}

The authors of \cite{Liu:2025frx} further conjectured that for a system with WH-MUB fiducial states but without WH-SICs, the $M_\alpha(| \psi \ket)$ is bounded by Eq.\ \eqref{bound} with $\Delta = 0$ and it is saturated uniquely by the WH-MUB fiducial states.

Table \ref{tab:magic_bounds} summarises the stabiliser and lattice-constructed maximal magic states in specific quantum systems.
\rev{For the 3-qubit and 1-qutrit systems, the displayed values are construction counts associated with $BW_{16}$ and $E_6$, respectively \cite{ohta_2025_15651358}.}

\section{Preliminaries: Symmetric lattices}

\begin{figure}[h]
 \centering
 \includegraphics[keepaspectratio, scale=0.15]
      {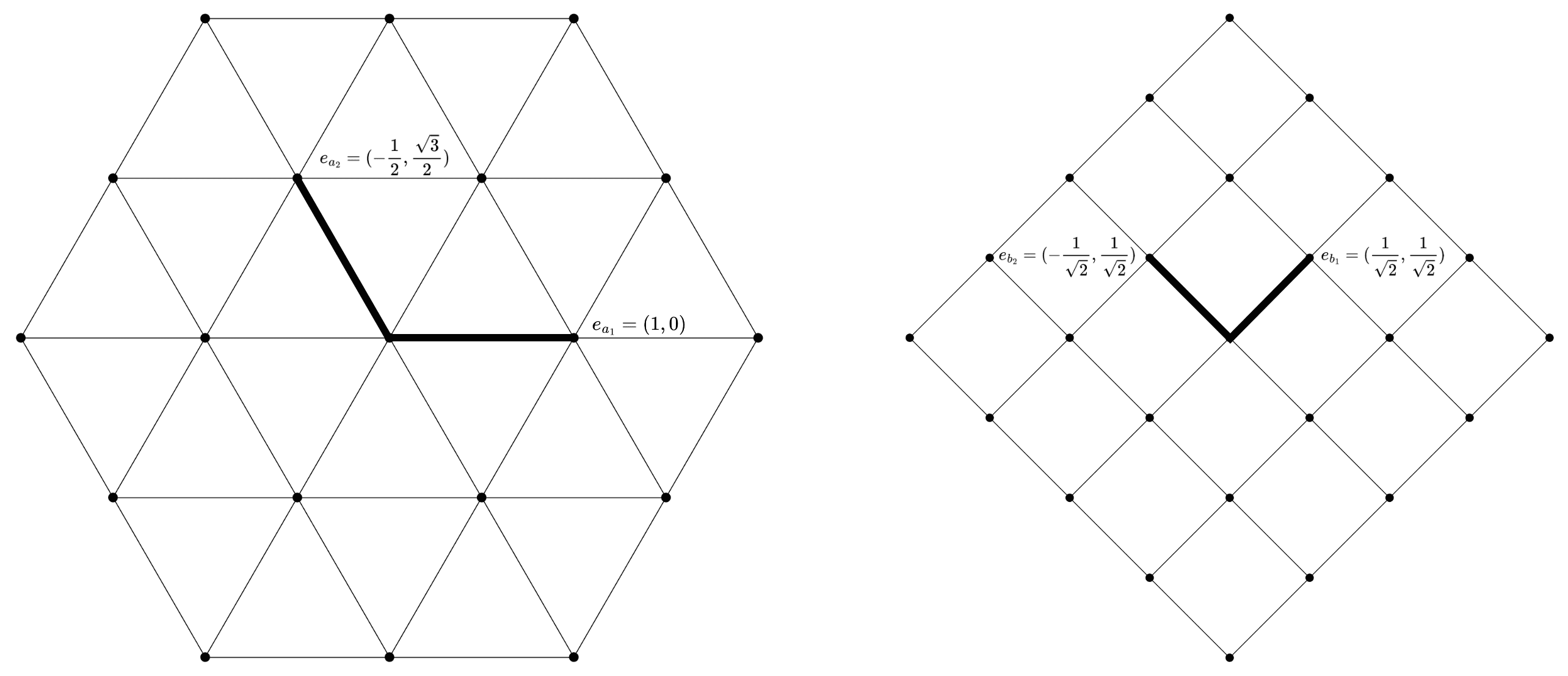}
 \caption{\small (a) Lattice points generated by $e_{a_1}$ and $e_{a_2}$ with norm up to 6 from the centor,(b) Lattice points generated by $e_{b_1}$ and $e_{b_2}$ with norm up to 8 from the centor.}
 \label{lattice}
\end{figure}

Intuitively, a lattice can be regarded as a set of regularly spaced points in space, which can be seen, for example, in crystal structures or periodic networks. 
Mathematically, a lattice $\Lambda$ in $\mathbb{R}^n$ is defined as the set 
\begin{equation}
\Lambda = \left\{ \vec{x} = a_1 \vec{v}_1 + \cdots + a_m \vec{v}_m \mid a_k \in \mathbb{Z} \right\}
\label{lattice_definition}
\end{equation}
where $\vec v_1, \ldots, \vec v_m$ are linearly independent vectors in $\mathbb{R}^n$ called the basis vectors of lattice $\Lambda$. 
Fig.\ \ref{lattice} shows parts of $A_2$ (left) and $D_2$ (right) lattices around the origin with their two basis vectors as examples.

In ${\mathbb R}^d$, each basis vector $\vec v_k$ is represented by an array of $n$ real numbers as 
$\vec v_k = (v_{k,1}, \ldots, v_{k,d})$.
The collection of $m$ basis vectors forms a $m \times d$ matrix, called the generator matrix of the lattice:
\begin{equation}
M\equiv\left(\begin{array}{c}
\vec{v}_1 \\
\vec{v}_2 \\
\vdots \\
\vec{v}_m
\end{array}\right)=\left(\begin{array}{cccc}
v_{1,1} & v_{1,2} & \cdots & v_{1 ,d} \\
\hline v_{2,1} & v_{2,2} & \cdots & v_{2 ,d} \\
\hline \vdots & \vdots & \ddots & \vdots \\
\hline 
v_{m, 1} & v_{m, 2} & \cdots & v_{m, d}
\end{array}\right) \,.
\end{equation}
If the lattice $\Lambda$ is generated by the generator matrix $M$ with integer coefficients as in \eqref{lattice}, we write $\Lambda=\textsf{Span}_{\mathbb Z} M$.
On the other hand, a lattice can be generated with a different set of coefficients, called the field.
For example, one may take
\be
\begin{aligned}
\mathbb{Z}[\omega] & =\left\{a+\omega b \mid a, b \in \mathbb{Z}, \omega=e^{2 \pi i / 3}\right\}\,, \\
\mathbb{Z}[i] & =\{a+i b \mid a, b \in \mathbb{Z}, i=\sqrt{-1}\}\,,
\end{aligned}
\ee
where $\mathbb{Z}[\omega]$ is known as the Eisenstein integers and plays an important role in the construction of the $E_6$ lattice.
If the lattice is generated by the generator matrix $M$ over the field $F$, we write
\bea
\Lambda &=& \textsf{Span}_F M 
\nonumber \\
&=& \{ \vec x = (a_1, \ldots, a_m) M  \mid a_k \in F \}
\eea

In the following discussion, we focus on sets of lattice points that share the same length, i.e., the distance from the origin. 
The vectors with the smallest, second-smallest, and third-smallest lengths are referred to as the first-shortest, second-shortest and third-shortest vectors, respectively, and so on.
For $F = \mathbb Z$, the square norm is given by
$\bra \vec x ,\vec x \ket = \vec{a} \cdot M M^T \cdot \vec a^T$
($ \vec a \in {\mathbb Z}^d$).
If the square norm of the $n$-th shortest vector is given by $\ell_n$, the collection of all $n$-th shortest vectors can be found by searching for all integer solutions of $ \vec{a} \cdot M M^T \cdot \vec a^T = \ell_n$ for $\vec a \in \mathbb Z^d$ in a given range of $a_k \in \{ a_k^{{\rm min},n}, a_k^{{\rm max},n} \}$.
A procedure to find such a search range is detailed in Appendix \ref{app:search}.

\section{$E_8$ lattice and two-qubit magic states}

The $E_8$ lattice points are generated as
\bea
\Lambda_{E_8} \,=\, \textsf{Span}_{\mathbb Z} M_{E_8} 
\,=\, \{ \vec x = (a_1, \ldots, a_8) M_{E_8}  \mid a_k \in {\mathbb Z} \}\,.
\eea
The generator matrix of $M_{E_8}$ is provided in Appendix \ref{app:generators}.
The square norm of the $E_8$ lattice points are given by $\bra \vec x, \vec x \ket = \vec a M_{E_8}  M_{E_8}^T \vec a^T$.
We list below the possible norms and the counts of those $E_8$ vectors \cite{conway2013sphere}:
\[
\begin{array}{r|r}
\bra \vec x, \vec x\ket & N(\bra \vec x, \vec x\ket) \\
\hline
0 & 1 \\
2 & 240 \\
4 &  2160 \\
6 & 6720 \\
8 & 17520 \\
10 & 30240 \\
\vdots & \vdots
\end{array}
\]


The complete list of the first-shortest vectors of $E_8$ can be obtained by searching for the integer solutions to $\vec a M_{E_8} M_{E_8}^T \vec a^T = 2$ in the finite range of $a_k \in {\mathbb Z}$ given in Appendix \ref{app:search}.
Examples of the first-shortest vectors are 
\bea
&& \vec x_1^{(1)} = (1,-1,0,0,0,0,0,0), \nonumber \\
&& \vec x_2^{(1)} = (1,0,0,0,0,0,0,1), \nonumber \\
&& \vec x_3^{(1)} = (-\tfrac{1}{2}, -\tfrac{1}{2}, -\tfrac{1}{2}, -\tfrac{1}{2}, -\tfrac{1}{2}, -\tfrac{1}{2}, \tfrac{1}{2}, \tfrac{1}{2})\,.  
\label{eg_vec1}
\eea
The first-shortest vectors of $E_8$ can be classified into two types.
The first type consists of vectors where six components are zero, and the remaining two components are either $1$ or $-1$, as demonstrated in $\vec{x}_1^{(1)}$ and $\vec{x}_1^{(2)}$ in the examples above. 
There are $\binom{8}{2} \times 4 = 112$ first-shortest vectors of this type.
The second type consists of the remaining 128 first-shortest vectors, where all components are either $\tfrac{1}{2}$ or $-\tfrac{1}{2}$, with each value appearing an even number of times, as exemplified by $\vec{x}_3^{(1)}$.

Using the map $(g \circ \varphi)(\vec x) = | \psi \ket$ defined in subsection \ref{sec:map}, we find the collection of pure two-qubit quantum states corresponding to the first-shortest vectors of the $E_8$ lattice. 
For instance, the above example lattice points are mapped to
\bea
\vec{x}_1^{(1)} & \xlongrightarrow{g \circ \varphi} & 
| \psi_1^{(1)} \ket = 
\tfrac{1}{\sqrt{2}} \left( | 00 \ket - | 01 \ket \right),
\nonumber \\
\vec{x}_2^{(1)} & \xlongrightarrow{g \circ \varphi} & 
| \psi_2^{(1)} \ket = 
\tfrac{1}{\sqrt{2}} \left( |00 \ket + i |11 \ket \right),
\nonumber \\
\vec{x}_3^{(1)} & \xlongrightarrow{g \circ \varphi} & 
| \psi_2^{(1)} \ket = 
\tfrac{1}{\sqrt{4}} \left( |00 \ket + |01 \ket -i |10 \ket - i |11 \ket \right).
\eea

We observe that for a given first-shortest vector, there exist three other first-shortest vectors that map to the same phase-equivalent quantum state.
These states are proportional to each other with the $\{ \pm1, \pm i\}$ phase factors. 
Consequently, the 240 first-shortest vectors yield 60 distinct two-qubit quantum states. The explicit forms of these 60 quantum states are provided in the supplemental material \cite{ohta_2025_15651358}.
For these quantum states, we compute the stabiliser Rényi entropy (SRE) using formulas \eqref{defM0} and \eqref{Xi2_nbits}. 
This analysis leads to the following key result:
\begin{itemize}
\item
\textit{All 60 two-qubit quantum states corresponding to the first-shortest vectors of the $E_8$ lattice are stabiliser states, satisfying $M_2(|\psi\rangle) = 0$.}
\end{itemize}


\noindent
It is well established that the two-qubit system has exactly 60 stabiliser groups, each corresponding to a unique stabiliser state. Thus, the above result can be rephrased as follows:
\textit{The complete set of 240 shortest vectors of the $E_8$ lattice, via the map $g \circ \varphi$, generates the full set of 60 stabiliser states of the two-qubit system.}

\medskip

Let us now turn our attention to the second-shortest vectors.
The number of second-shortest vectors in the $E_8$ lattice (with length $\ell_2 = 4$) is known to be 2,160. 
The complete list of the second-shortest vectors of $E_8$ can be obtained by searching for the integer solutions to $\vec a M_{E_8} M_{E_8}^T \vec a^T = 4$ in the range of $a_k \in {\mathbb Z}$ given in Appendix \ref{app:search}.
These second-shortest vectors are mapped to pure quantum states of the two-qubit system via the function $g \circ \varphi$. 
As noted above, we find that four of the second-shortest vectors map to the same phase-equivalent quantum state. Consequently, the 2,160 second-shortest vectors correspond to 540 distinct quantum states.
Examples of the second-shortest vectors and the corresponding quantum states are
\bea
\vec{x}_1^{(2)} = (0,0,0,1,0,1,1,1) &\xlongrightarrow{g \circ \varphi}&
| \psi_1^{(2)} \ket = \tfrac{1}{2} \left( |01 \ket + |10 \ket + (1 - i)|11 \ket \right), 
\nonumber \\
\vec{x}_2^{(2)} = \left(-\tfrac{1}{2}, \tfrac{3}{2}, \tfrac{1}{2},-\tfrac{1}{2}, \tfrac{1}{2},-\tfrac{1}{2},-\tfrac{1}{2},-\tfrac{1}{2}\right)
&\xlongrightarrow{g \circ \varphi}&
| \psi_2^{(2)} \ket = 
\tfrac{1}{\sqrt{6}} \left( | 00 \ket -(2+i) |01 \ket - |10 \ket -i|11 \ket \right), 
\nonumber \\
\vec{x}_3^{(2)} = \left(-\tfrac{1}{2}, \tfrac{1}{2}, \tfrac{1}{2}, \tfrac{1}{2}, \tfrac{3}{2}, \tfrac{1}{2},-\tfrac{1}{2},-\tfrac{1}{2}\right)
&\xlongrightarrow{g \circ \varphi}&
| \psi_3^{(2)} \ket = 
\tfrac{1}{\sqrt{6}} \left( (-2+i)| 00 \ket +i |01 \ket + |10 \ket + |11 \ket \right).
\nonumber \\
\label{2bit_2nd_example}
\eea
%
%

We then compute the SRE for those 540 quantum states.
We observed that out of the 540 states, 60 states have $M_2(| \psi \ket) = 0$.
These states turn out to be phase-equivalent to the stabiliser states found in the analysis of the first-shortest vectors. 
The remaining 480 states have $M_2(| \psi \ket) = - \log \tfrac{7}{16} \simeq 1.193$ (i.e., $\Xi_2(| \psi \ket) = \tfrac{7}{16}$).
As discussed in section \ref{sec:magic}, this coincides with the maximum of the SRE in the two-qubit system \cite{Liu:2025frx}. 
It has been known that the two-qubit system has exactly 480 maximal magic states with $M_2(| \psi \ket) = - \log \tfrac{7}{16}$ \cite{Liu:2025frx}.
We summarise the key result of this analysis as follows:
\begin{itemize}
\item
\textit{All 540 two-qubit quantum states corresponding to the second-shortest vectors of the $E_8$ lattice split into the complete 60 stabiliser states with $M_2(|\psi\rangle) = 0$ and the complete 480 maximal magic states with $M_2(|\psi\rangle) = -\log \tfrac{7}{16}$.}
\end{itemize}


\noindent
For instance, three example states in Eq.\ \eqref{2bit_2nd_example} are the maximal magic states.

\begin{figure}[t!]
 \centering
 \includegraphics[keepaspectratio, scale=0.48]
      {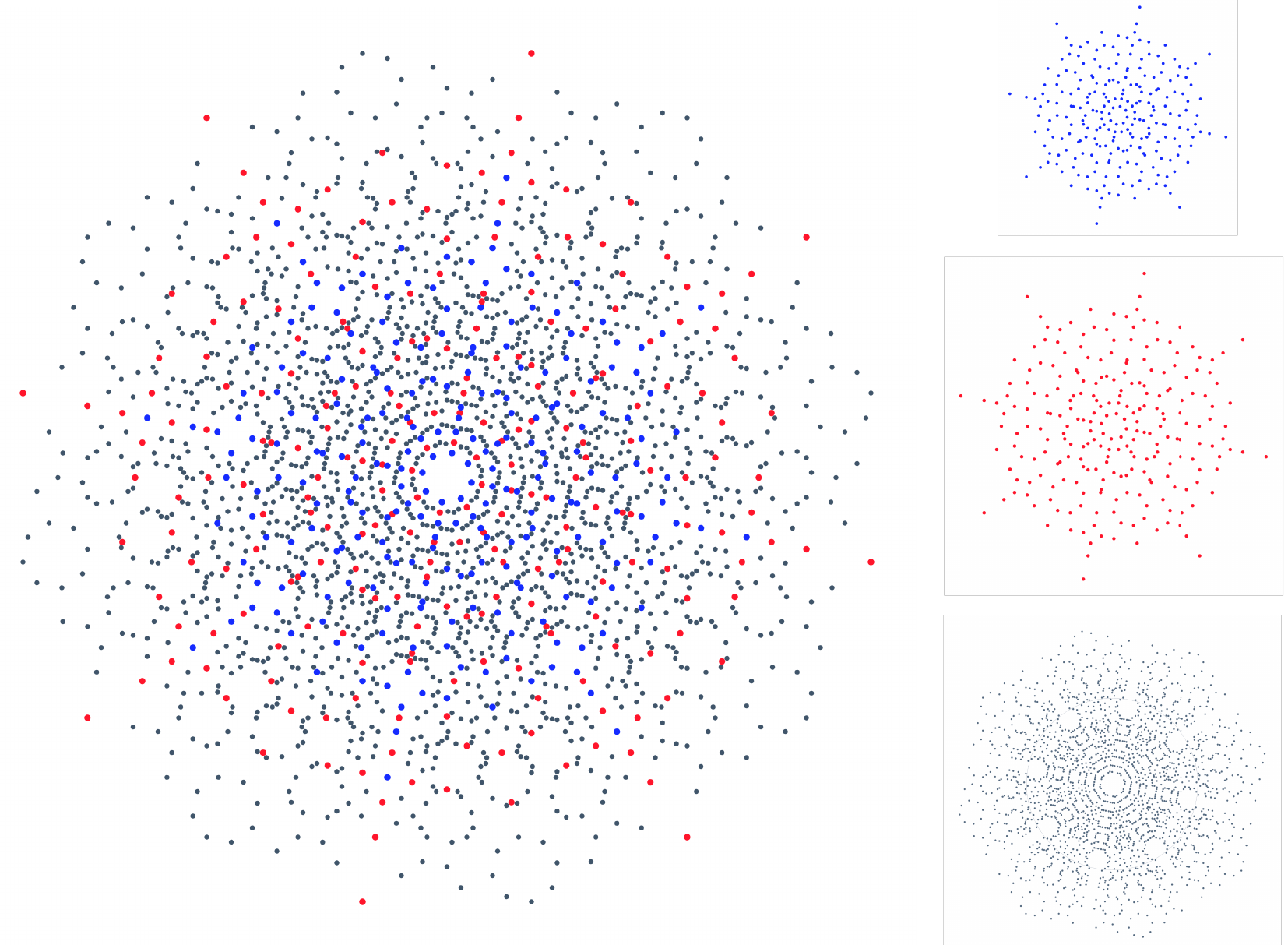}
 \caption{\small Two-dimensional projection of the first- and second-shortest vectors of the $E_8$ lattice obtained by the projection matrix \eqref{E8proj}.
 Blue points represent the first-shortest vectors, while red and grey points denote the second-shortest vectors associated with the stabiliser and maximal magic states, respectively.
 On the left, all three sets of points are displayed on the same plane for comparison; on the right, they are shown separately but at the same scale. 
}
 \label{E8_12vectors}
\end{figure}

Fig.\ \ref{E8_12vectors} shows orthogonal projections of the first- and second-shortest vectors of the $E_8$ lattice using the projection matrix ($\mathbb{R}^8 \to \mathbb{R}^2$) 
\be
\mathcal{P} = \frac{1}{2}
\begin{pmatrix}
1 & \cos\left(\frac{\pi}{8}\right) & \cos\left(\frac{2\pi}{8}\right) & \cdots & \cos\left(\frac{7\pi}{8}\right) \\
0 & \sin\left(\frac{\pi}{8}\right) & \sin\left(\frac{2\pi}{8}\right) & \cdots & \sin\left(\frac{7\pi}{8}\right)
\end{pmatrix}\,.
\label{E8proj}
\ee
In the figure, the red points represent the first-shortest vectors, while the blue and grey points correspond to the second-shortest vectors associated with the stabiliser and maximal magic states, respectively.
Due to the projection from high-dimensional space to a lower-dimensional plane, a single dot in the plot may correspond to multiple lattice vectors.
On the left, all three sets of points are displayed on the same plane for comparison; on the right, they are shown separately but at the same scale.
Notably, the distribution of second-shortest vectors corresponding to stabiliser states (red) closely resembles that of the first-shortest vectors (blue).

A relationship between magic and entanglement is explored in Ref.\ \cite{Liu:2025frx}.
Using concurrence $\mathcal{C}_{AB}$ \cite{Wootters:1997id,Hill:1997pfa} to quantify the entanglement between qubits $A$ and $B$, it is found that 36 stabiliser states exhibit no entanglement $(\mathcal{C}_{AB} = 0)$, corresponding to separable states.
The remaining 24 stabiliser states are maximally entangled, with $\mathcal{C}_{AB} = 1$.
In contrast, among the 480 maximal magic states, 192 states have $\mathcal{C}_{AB} = \tfrac{1}{2}$, while the remaining 288 exhibit $\mathcal{C}_{AB} = \tfrac{1}{\sqrt{2}}$.

\renewcommand{\arraystretch}{1.1} 
\begin{table}[t!]
\begin{center}
\begin{tabular}{|c||c|c|c|c|c||r|}
\hline
$\bra x,x \ket$ &
$M_2^{\rm min}=0$ &
$\log \left(\frac{27}{19}\right)$ &
$\log \left(\frac{9}{5}\right)$ &
$\log \left(\frac{256}{139}\right)$ &
$M_2^{\rm max}=\log \left(\frac{16}{7}\right)$ &
${\rm Total} \times 4$  
\\
\hline
\hline
2 & 60 & 0 & 0 & 0 & 0 & 240 \\
\hline
4 & 60 & 0 & 0 & 0 & 480 & 2160 \\
\hline
6 & 0 & 960 & 720 & 0 & 0 & 6720\\
\hline
8 & 60 & 0 & 0 & 3840 & 480 & 17520\\
\hline
\end{tabular}
\caption{\small 
The stabiliser Rényi entropy values, $M_2(|\psi\rangle)$, for the two-qubit states derived from the first-, second-, third-, and fourth-shortest vectors of the $E_8$ lattice. The numbers in the table represent the count of quantum states within each category.
\label{two-qubit_summary}
}
\end{center}
\end{table}

It is natural to ask how the above analysis extends to the third- and fourth-shortest vectors of the $E_8$ lattice.
Applying the same methodology, we find that the 6,720 third-shortest vectors yield 1,680 distinct quantum states. 
These split into two groups: 960 states with $M_2(| \psi \rangle) = - \log \tfrac{19}{27}$ and 720 states with $M_2(| \psi \rangle) = - \log \tfrac{5}{9}$.
Similarly, the 17,520 fourth-shortest vectors generate 4,380 distinct quantum states, which fall into three categories based on their SRE values:
240 states with $M_2(| \psi \rangle) = 0$ (stabiliser states),
1,920 states with $M_2(| \psi \rangle) = - \log \tfrac{7}{16}$ (maximal magic states),
and the remaining 15,360 states with the intermediate value $M_2(| \psi \rangle) = - \log \tfrac{139}{256}$.
A summary of these results is presented in Table~\ref{two-qubit_summary}.

\section{$BW_{16}$ lattice and 3-qubit magic states}
\label{sec:3-qubit}

The Barnes-Wall lattice, $BW_{16}$, was first introduced by Barnes and Wall in 1959 \cite{Barnes_Wall_1959}. Since then, it has been independently rediscovered by several researchers (see Ref.\ \cite{conway2013sphere} for details). 
The lattice points of $BW_{16}$ are generated by
\be
{\Lambda}_{BW_{16}}\,=\,\textsf{Span}_{\mathbb{Z}}\, M_{BW_{16}}
\,=\, \{ \vec x = \vec{a} \cdot M_{BW_{16}} \mid a_k \in {\mathbb Z} \}\,.
\ee
The generator matrix, $M_{BW_{16}}$, is given in Appendix \ref{app:generators}.
The first-, second-, third- and fourth-shortest vectors of the $BW_{16}$ lattice have the square norm $\bra \vec x, \vec x\ket$ of 4, 6, 8 and 10, respectively. 
A partial list for the number of those vectors is given below \cite{conway2013sphere}:
\[
\begin{array}{r|r}
\bra \vec x, \vec x\ket & N(\bra \vec x, \vec x\ket) \\
\hline
0 & 1 \\
2 & 0 \\
4 & 4320 \\
6 & 61440 \\
8 & 522720 \\
10 & 2211840 \\
\vdots & \vdots
\end{array}
\]

As in the $E_8$ case, we generate all 4,320 first-shortest vectors by searching for the integer solutions to $\vec a M_{BW_{16}} M_{BW_{16}}^T \vec{a}^T = 4$ in the range of $a_k$ given in Appendix \ref{app:search}.
Next, we map the obtained first-shortest vectors to three-qubit quantum states with the $2^3 = 8$ computational basis states.
For example, two first-shortest vectors 
\bea
\vec x_1^{(1)} &=& \left(\tfrac{1}{2}, \tfrac{1}{2}, \tfrac{1}{2}, \tfrac{1}{2}, \tfrac{1}{2}, \tfrac{1}{2}, \tfrac{1}{2}, \tfrac{1}{2}, \tfrac{1}{2}, \tfrac{1}{2}, \tfrac{1}{2}, \tfrac{1}{2},-\tfrac{1}{2},-\tfrac{1}{2},-\tfrac{1}{2},-\tfrac{1}{2}\right), 
\nonumber \\
\vec x_2^{(1)} &=& (0,1,1,0,0,0,0,0,0,0,0,0,0,1,1,0)\,,
\eea
are mapped to the following three-qubit states 
\bea
\vec x_1^{(1)} &\xlongrightarrow{g \circ \varphi}& | \psi_1^{(1)} \ket \,=\, \tfrac{1}{2 \sqrt{2}} \left( | 0 \ket - i | 1 \ket \right) \otimes
\left( | 00 \ket + | 01 \ket + | 10 \ket + | 11 \ket
\right),
\nonumber \\
\vec x_2^{(1)} &\xlongrightarrow{g \circ \varphi}&  | \psi_2^{(1)} \ket \,=\, \tfrac{1}{2} \left( | 0 \ket + i | 1 \ket \right) \otimes
\left( | 01 \ket + | 10 \ket \right),
\eea
respectively. 

Due to the symmetry of the $BW_{16}$ lattice, we find that always four of the first-shortest vectors map to the same phase-equivalent quantum states, which are related to each other by the $\{ \pm1, \pm i\}$ phases. 
Consequently, the 4,320 shortest vectors yield 1,080 distinct quantum states. 
We then calculate the stabiliser Rényi entropy for these 1,080 states and find that all of them satisfy $M_2(|\psi\rangle) = 0$. 
Notably, it is known that the three-qubit system possesses exactly 1,080 stabiliser states with $M_2(|\psi\rangle) = 0$.
We are led to the following conclusion:
\begin{itemize}
\item
\textit{The complete set of 4,320 shortest vectors of the $BW_{16}$ lattice generates the full set of 1,080 stabiliser states of the three-qubit system.}
\end{itemize}


Next, we turn our attention to the second-shortest vectors of the $BW_{16}$ lattice. The complete set of 61,440 second-shortest vectors can be obtained by collecting all integer solutions to $\vec{a} M_{BW_{16}} M_{BW_{16}}^T \vec{a}^T = 6$, with $a_k$ constrained as specified in Appendix \ref{app:search}.
Using the function $g \circ \varphi$, these second-shortest vectors are mapped to three-qubit quantum states.  
For example, the second-shortest vectors 
\bea
\vec x_1^{(2)} &=& \left(\tfrac{1}{2}, \tfrac{1}{2}, \tfrac{1}{2}, \tfrac{1}{2}, \tfrac{1}{2}, \tfrac{3}{2},-\tfrac{1}{2}, \tfrac{1}{2}, \tfrac{1}{2}, \tfrac{1}{2}, \tfrac{1}{2}, \tfrac{1}{2}, \tfrac{1}{2},-\tfrac{1}{2},-\tfrac{1}{2}, \tfrac{1}{2}\right)\,, 
\nonumber \\
\vec x_2^{(2)}  &=& (0,0,1,0,0,1,0,0,0,0,1,0,1,0,1,1)\,,
\eea
are mapped to the following states 
\bea
\vec x_1^{(2)} &\xlongrightarrow{g \circ \varphi}& | \psi_1^{(2)} \ket \,=\, \tfrac{1}{2 \sqrt{3}} \left(  
| 000 \ket + | 001 \ket + | 010 \ket + | 011 \ket + | 100 \ket + (1 - 2 i) | 101 \ket - | 110 \ket + | 111 \ket 
\right),
\nonumber \\
\vec x_2^{(2)} &\xlongrightarrow{g \circ \varphi}&  | \psi_2^{(2)} \ket \,=\, \tfrac{1}{\sqrt{6}} \left( (1-i) | 010 \ket + | 100 \ket -i | 101 \ket + | 110 \ket + | 111 \ket \right),
\eea
respectively.

Due to the lattice symmetries, the 61,440 second-shortest vectors yield 15,360 distinctive quantum states.
By computing the stabiliser R{\' e}nyi entropy, we find all these states have $M_2(| \psi \ket) = - \log \tfrac{2}{9}.$
From the discussion in section \ref{sec:magic}, we notice that the observed value saturates the maximum of the SRE for the three-qubit system.   
We are led to the following conclusion:
\begin{itemize}
\item
\textit{The full set of 61,440 second-shortest vectors of the $BW_{16}$ lattice generates 15,360 three-qubit states with the maximal magic, $M_2(| \psi \ket) = - \log \tfrac{2}{9}$.}
\end{itemize}

\begin{revision}
The appearance of the three-qubit WH-SIC can be understood from the special symmetry of this shell.  The three-qubit Clifford generators preserve the complex form of $BW_{16}$, and one second-shell representative has squared overlap $1/9$ with every non-identity Pauli operator; hence every ray in its Clifford orbit is a Hoggar SIC fiducial \cite{NebeRainsSloane2001,kliuchnikov2024stabilizeroperatorsbarneswalllattices,SzymusiakSlomczynski2016}.  Its projective stabiliser is $\operatorname{PSU}(3,3)$ of order $6{,}048$, so orbit--stabiliser gives $15{,}360$ rays, exactly the second-shell count $61{,}440/4$ after the four Gaussian unit phases are removed \cite{Mastel:2023uhk,conway2013sphere,ohta_2025_15651358}.  Thus the Clifford orbit of Hoggar fiducials exhausts the second shell.  This exceptional symmetry is independently reflected in Zhu's classification: the 64 Hoggar lines generated by a fiducial form are, up to unitary equivalence, one of only three SICs whose symmetry group is doubly transitive \cite{Zhu2015SuperSIC}.
\end{revision}


To the best of our knowledge, neither the total number of maximal magic states nor their explicit forms have been determined for the three-qubit system. 
Our findings provide explicit expressions for 15,360 maximal magic states within this system. 
The complete list of these states is included in the supplemental material \cite{ohta_2025_15651358}.

Given the highly symmetric structure of the $BW_{16}$ lattice and the role of WH-SIC as maximal magic states, we propose the following conjecture: the identified 15,360 states represent the complete set of maximal magic states for the three-qubit system.
This conjecture can be succinctly stated as:
\begin{description}
\item[Conjecture 1:]
\textit{The total number of maximal magic states in the three-qubit system is 15,360.}
\end{description}

\renewcommand{\arraystretch}{1.1} 
\begin{table}[t!]
\begin{center}
\begin{tabular}{|c||c|c|c|c||r|}
\hline
$\bra \vec x, \vec x \ket$ &
$M_2^{\rm min}= 0$ &
$\log \left(\frac{16}{7}\right)$ &
$\log \left(\frac{32}{11}\right)$ &
$M_2^{\rm max}=\log \left(\frac{9}{2}\right)$ &
${\rm Total} \times 4$  
\\
\hline
\hline
4 & 1080 & 0 & 0 & 0 & 4320 \\
\hline
6 & 0 & 0 & 0 & 15360 & 61440 \\
\hline
8 & 1080 & 60480 & 69120 & 0 & 522720 \\
\hline
\end{tabular}
\caption{\small 
The stabiliser Rényi entropy values, $M_2(|\psi\rangle)$, for three-qubit states derived from selected vectors of the $BW_{16}$ lattice. The table shows the number of quantum states corresponding to each entropy value, grouped by $\bra \vec x, \vec x \ket$.
\label{three-qubit_summary_alt}
}
\end{center}
\end{table}

We also investigate the magic states obtained from $BW_{16}$ lattice points with lengths exceeding those of the second-shortest vectors. 
For $n > 2$, the number of the $n$-th shortest vectors in the $BW_{16}$ lattice becomes significantly large: the counts for the third and fourth-shortest vectors are 522,720 and 2,211,840, respectively. 
Consequently, our exploration is restricted to the third-shortest vectors.

After resolving phase equivalence, the 522,720 third-shortest vectors produce 130,680 distinct three-qubit states. 
These states can be grouped into three categories based on their SRE values. 
Among them, 1,080 states are identified as stabiliser states, characterised by $M_2(|\psi\rangle) = 0$. 
A total of 60,480 states have an SRE value of $M_2(|\psi\rangle) = \log \left( \frac{16}{7} \right)$, while the remaining 69,120 states are found to have $M_2(|\psi\rangle) = \log \left( \frac{32}{11} \right)$.
The resulting magic values and the corresponding state counts are summarised in Table \ref{three-qubit_summary_alt}.


\subsection{Entanglement of three-qubit extremal magic states}
\label{sec:entangle}

Since the first- and second-shortest vectors of the $BW_{16}$ lattice provide closed-form expressions for the stabiliser and maximal magic states, respectively, it is straightforward to analyse their entanglement structures.  
In the system of three qubits $A$, $B$ and $C$, one can define three types of entanglement: 
\begin{description}
\item[one-to-one:] the entanglement between one qubit and another, quantified by the one-to-one concurrence, e.g., ${\cal C}_{AB}$.
\item[one-to-other:] the entanglement between one qubit and the rest of the system, quantified by the one-to-other concurrence, e.g., ${\cal C}_{A(BC)}$.
\item[genuine three-partite:] the entanglement among all three qubits, measured by the area of the concurrence triangle: ${\cal F}_3$.
\end{description}
The precise mathematical definitions of these entanglement measures are given in Appendix~\ref{app:entangle}.

\begin{figure}[t!]
 \centering
\includegraphics[keepaspectratio, scale=0.15]
      {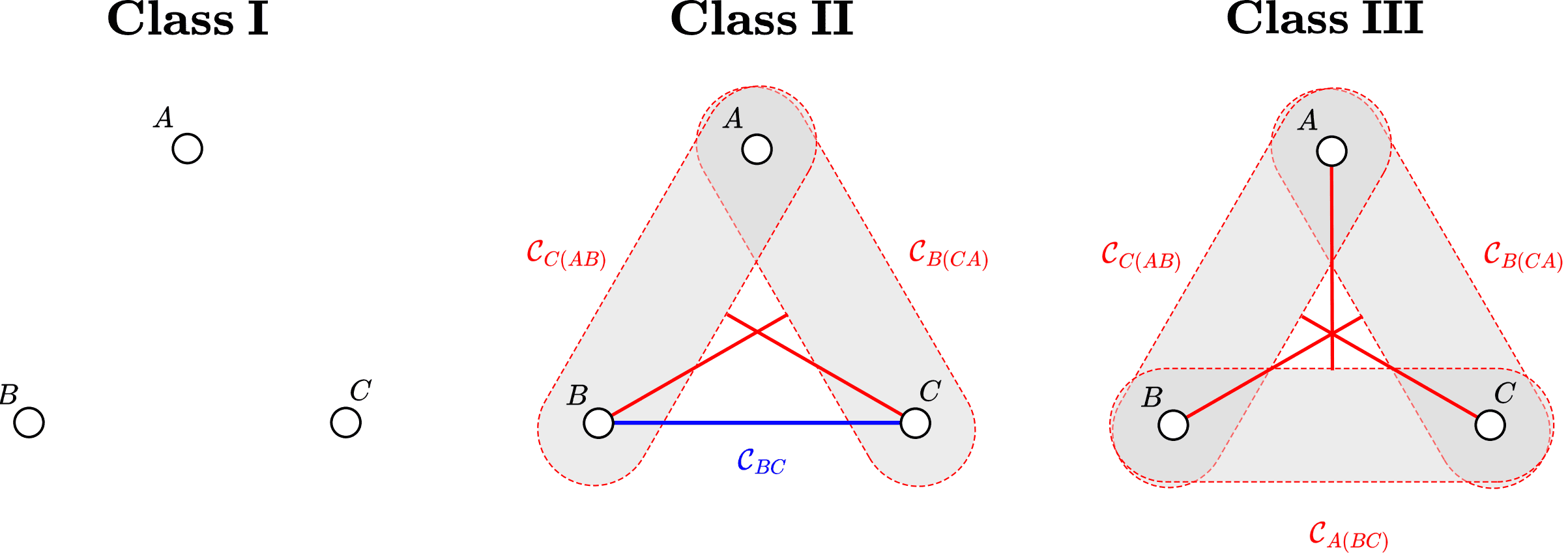}
 \caption{\small The figure of the class of the entanglement for the first-shortest vectors of $BW_{16}$. Blue lines represent one-to-one entanglement, red lines indicate one-to-other entanglement.}
 \label{entangle}
\end{figure}

We start by discussing the stabiliser states ($M_2(| \psi \ket) = 0$) in the three-qubit system.
We observe that the entanglement structures of the three-qubit stabiliser states are categorised into three distinct classes, shown in Fig.\ \ref{entangle}.
In Class-I, the states are fully separable of the form $| \psi \ket_{\rm fs} = | a \ket \otimes | b \ket \otimes | c \ket$.
All three types of entanglement vanish in this class, i.e.,
${\cal C}_{ij}={\cal C}_{i(jk)}={\cal F}_3=0$.
We find 216 stabiliser states belong to Class-I.

In Class-II, the states are characterised such that one qubit is separable while the other two are maximally entangled.  
Labelling the separated qubit by $A$, we have ${\cal C}_{BC} = {\cal C}_{B(AC)} = {\cal C}_{C(AB)} = 1$.
All other concurrence measures, as well as the genuine three-particle measure, vanish in this class. 
In Class-II, we find 432 stabiliser states. 

The states in Class-III are GHZ-type, in which all three one-to-other concurrences are maximal (${\cal C}_{i(jk)} = 1$), while all three one-to-one concurrences vanish (${\cal C}_{ij} = 0$).  
The genuine three-partite entanglement is also maximal in this case, ${\cal F}_3 = 1$.
We observe 432 stabiliser states in Class-III.


\begin{figure}[t!]
 \centering
\includegraphics[keepaspectratio, scale=0.15]
      {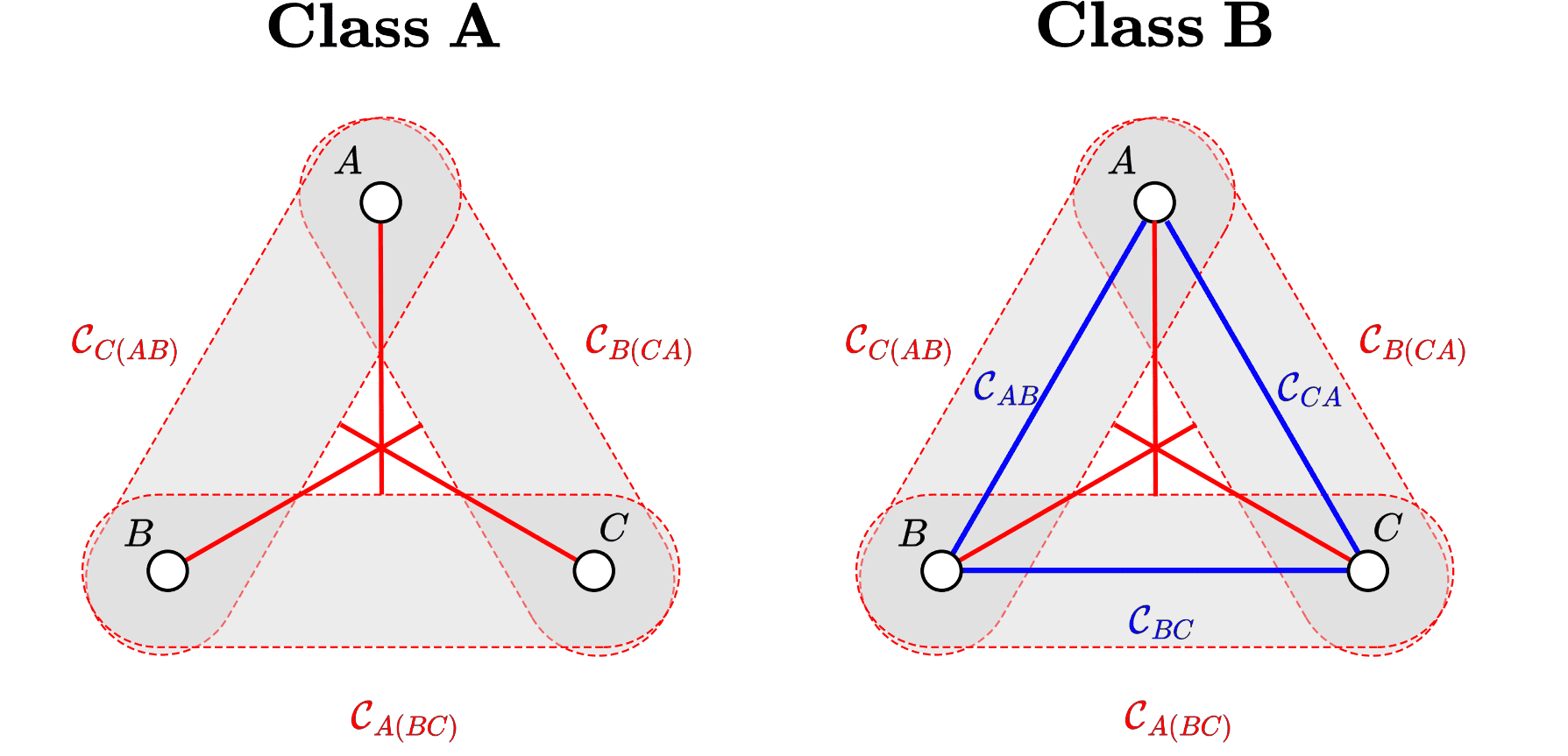}
 \caption{\small The figure of the the class of the entanglment for the second-shortest vectors of $BW_{16}$ (Maximally magic states). Blue lines represent one-one entanglement, red lines indicate one-two entanglement.}
 \label{entangle_max}
\end{figure}

We now focus on the maximal magic states of the three-qubit system. 
Notably, all 15,360 maximal magic states exhibit identical values for the one-to-other and genuine three-partite entanglement measures:
\be
\mathcal{C}_{A(BC)} = \mathcal{C}_{B(CA)} = \mathcal{C}_{C(AB)} = \frac{\sqrt{6}}{3}, \quad
\mathcal{F}_3 = \frac{2}{3}.
\ee
However, for the one-to-one entanglement measures, the states can be divided into two distinct classes (see Fig.\ \ref{entangle_max}).
In Class-A, all one-to-one entanglement measures vanish, $\mathcal{C}_{ij} = 0$.
Out of 15360 maximal magic states, 1536 of them belong to Class-A.
In Class-B, all one-to-one concurrence measures take the same nonzero value, $\mathcal{C}_{ij} = \frac{\sqrt{2}}{3}$. 
Class-B comprises the remaining 13824 maximal magic states. 

\begin{figure}[t!]
 \centering
\includegraphics[keepaspectratio, scale=0.45]
      {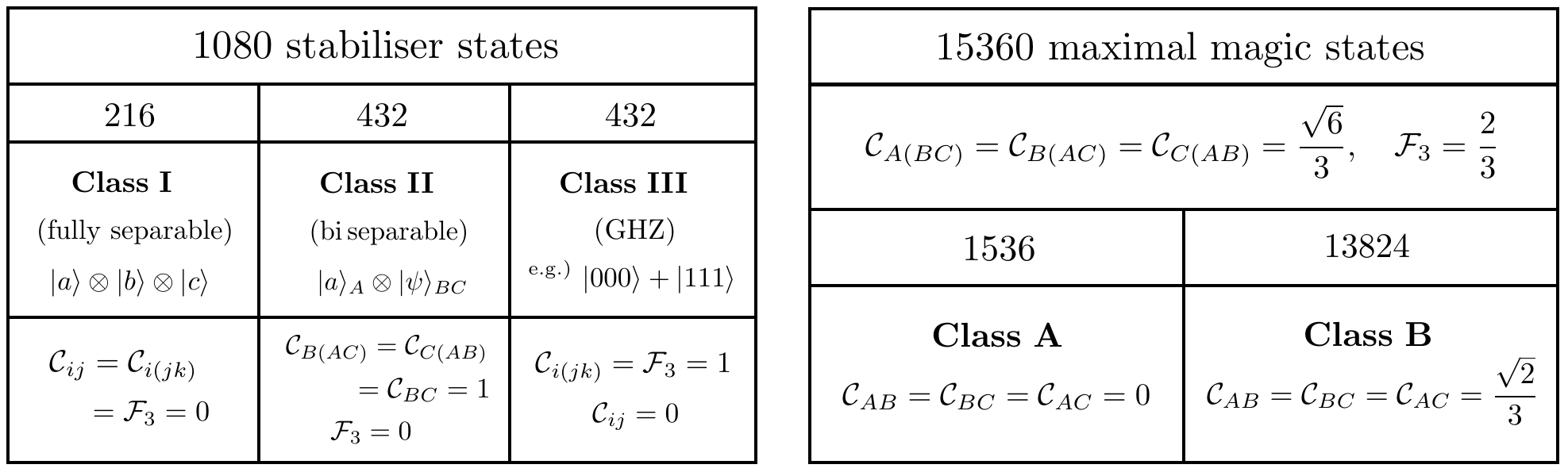}
 \caption{\small Entanglement characteristics of the stabiliser and maximal magic states in the three-qubit system.}
 \label{magic-entangle}
\end{figure}

Fig.\ \ref{magic-entangle} summarises the entanglement classification for the stabiliser and maximal magic states in the three-qubit system.


\section{$E_6$ lattice and 1-qutrit magic states}

The $E_6$ lattice can be defined in a three-dimensional complex space. 
In this construction, the $E_6$ lattice points are generated by
\cite{conway2013sphere}
\be
\Lambda_{E_6} \,=\, \textsf{Span}_{\mathbb{Z}[\omega]} M_{E_6}
\,=\, \{ \vec{c} = (\beta_1, \beta_2, \beta_3) M_{E_6} \mid 
\beta_k \in {\mathbb Z}[\omega] \}\,,
\ee
where the generator matrix is given by
\begin{equation}
M_{E_6}=\left(\begin{array}{lll}
\theta & 0 & 0 \\
0 & \theta & 0 \\
1 & 1 & 1
\end{array}\right)
\label{E6_gen}
\end{equation}
with $\omega = e^{2 \pi i /3}$ and $\theta \equiv \omega - \omega^2 =i\sqrt{3}$.
The components of the coefficient vector $\vec \beta$ is drawn from the set of Einstein integers 
$\mathbb{Z}[\omega]=\{a+\omega b \mid a, b \in \mathbb{Z}, \omega=e^{2\pi i/3}\}$.
The norm of the $E_6$ lattice vectors is defined by $\bra \vec c, \vec c \ket = |c_1|^2 + |c_2|^2 + |c_3|^2 = \vec \beta M_{E_6} M_{E_6}^\dagger \vec{\beta}^\dagger$.
A partial list of the $E_6$ square norms and the counts of corresponding vectors is shown below \cite{conway2013sphere}:\footnote{To match this list with the one that is conventionally known as $E_6$, we need to multiply by a factor of $\alpha = \sqrt{\tfrac{2}{3}}$ to the vectors generated by the matrix \eqref{E6_gen}.} 
\[
\begin{array}{r|r}
\bra \vec c, \vec c \ket & N(\bra \vec c, \vec c \ket) \\
\hline
0 & 1 \\
3 & 72 \\
6 & 270 \\
9 & 720 \\
12 & 936 \\
15 & 2160 \\
\vdots & \vdots
\end{array}
\]


We construct all 72 first-shortest vectors and 270 second-shortest vectors by collecting all solutions to $\vec \beta M_{E_6} M_{E_6}^\dagger \vec{\beta}^\dagger = 2$ and 4, respectively, in the domain $\beta_k \in a_k + b_k \omega$, $a_k,b_k \in {\mathbb Z}$.
The search ranges for $a_k,b_k \in {\mathbb Z}$ are provided in Appendix \ref{app:search}.


Using the map $g$ defined in subsection \ref{sec:map}, we obtain the collection of single-qutrit quantum states generated by the first-shortest vectors of $E_6$.
Due to the symmetries of the $E_6$ lattice in the complex space, always six $n$-th shortest vectors are mapped to six phase-equivalent quantum states, related to each other with the $\{\pm1, \pm \omega, \pm \omega^2 \}$ phase factors. 
Consequently, the set of 72 first-shortest vectors of $E_6$ generates 12 distinct quantum states.
Examples of the first-shortest vectors and all 12 corresponding quantum states are explicitly given in subsection \ref{sec:E6stab}.

As in the previous study, we calculate the SRE using the formulae \eqref{defM0} and \eqref{Xi} with summation taken over all 9 elements of the phase-quotiented Weyl-Heisenberg group $\mathcal{W}(1,3)$ 
(see subsection \ref{sec:E6stab} for details).
Through this exercise, we find that all 12 single-qutrit quantum states obtained from the first-shortest vectors of $E_6$ have $M_2(| \psi \ket) =0$.
Incidentally, it is known that there are a total of 12 stabiliser states in the single-qutrit system.
We conclude:

\begin{itemize}
\item
\textit{The full set of 72 shortest vectors of the $E_{6}$ lattice generates the full set of 12 stabiliser states of the single-qutrit system.}
\end{itemize}

Next, we transform the second-shortest vectors into quantum states using the function $g$.
Again, due to the symmetry of the $E_6$ lattice, always six lattice points are mapped to the same phase-equivalent states.
The set of 270 second-shortest vectors, therefore, generates 45 distinct quantum states of a 1-qutrit. 
Two examples of second-shortest vectors and the corresponding quantum states are:
\bea
\vec c_1^{\,(2)} = (\theta, \theta, 0) 
&\xlongrightarrow{g}& |\psi_1^{(2)} \ket = \tfrac{1}{\sqrt{2}} \left( | 1 \ket + | 2 \ket \right),
\nonumber \\
\vec c_2^{\,(2)} = (0, \omega^2 \theta, -\omega \theta) 
&\xlongrightarrow{g}& |\psi_2^{(2)} \ket = \tfrac{1}{\sqrt{2}} \left( e^{2 \pi i / 3} | 2 \ket - | 3 \ket \right).
\label{qutrit_example}
\eea

Computing the SRE for those quantum states, we find all 45 states give the same value $M_2(| \psi \ket) = 1$.  
From the discussion in section \ref{sec:magic}, we note that this value saturates the upper bound of the SRE for the single-qutrit system. 
We are led to the following conclusion:

\begin{itemize}
\item
\textit{The full set of 270 second-shortest vectors of the $E_{6}$ lattice generates 45 maximal magic states of the single-qutrit system.}
\end{itemize}

\begin{revision}
The number 45 is the size of the maximal-magic subset selected by the second shell of $E_6$, not the total number of one-qutrit maximisers.  Qutrit WH-SIC fiducials form a continuous one-parameter family, so our finite count is a statement about this $E_6$ shell rather than a classification of every one-qutrit maximal magic state \cite{Appleby2005,TabiaAppleby2013}.
\end{revision}

It is intriguing to explore the magic states generated by the third-, fourth-, and fifth-shortest vectors of the $E_6$ lattice through the map $g$.

Through direct calculations, we find that the 720 third-shortest vectors yield 120 distinct single-qutrit states. These states are divided into two groups: one comprising 12 stabiliser states ($M_2(| \psi \rangle) = 0$) and the other containing 108 states with $M_2(| \psi \rangle) = \log \left( \frac{625}{401} \right)$.

Similarly, the 936 fourth-shortest vectors generate 156 distinct states. These consist of 12 stabilizer states and 144 states with $M_2(| \psi \rangle) = \log \left( \frac{32}{17} \right)$.

Finally, we analyse the states derived from 1,260 fifth-shortest vectors, which produce 210 distinct states. These are categorized into two groups: 216 states with $M_2(| \psi \rangle) = \log \left( \frac{625}{401} \right)$ and 144 states with $M_2(| \psi \rangle) = \log \left( \frac{625}{353} \right)$.

The resulting magic values and the corresponding counts of states are summarised in Table \ref{qutrit_summary}.

\renewcommand{\arraystretch}{1.1} 
\begin{table}[t!]
\begin{center}
\begin{tabular}{|c||c|c|c|c|c|c||r|}
\hline
$\bra \vec c, \vec c \ket$ 
& $M_2^{\rm min}= 0$ 
& $\log\left( \tfrac{625}{401} \right)$ 
& $\log\left( \tfrac{81}{49} \right) $
& $\log\left( \tfrac{625}{353} \right)$ 
& $\log\left( \tfrac{32}{17} \right) $
& $M_2^{\rm max}=1$ 
& ${\rm Total} \times 6$
\\
\hline\hline
3  & 12  & 0    & 0   & 0   & 0   & 0 & 72   \\
\hline
6  & 0   & 0    & 0   & 0   & 0   & 45 & 270 \\
\hline
9  & 12  & 0    & 108 & 0   & 0   & 0  & 720 \\
\hline
12  & 12  & 0    & 0   & 0   & 144 & 0  & 936 \\
\hline
15 & 0   & 216 & 0   & 144 & 0   & 0 & 1260  \\
\hline
\end{tabular}
\caption{\small
The stabiliser Rényi entropy values, $M_2(|\psi\rangle)$, for the single-qutrit states derived from the first- to fifth- shortest vectors of the $E_6$ lattice. The numbers in the table represent the count of quantum states within each category. 
\label{qutrit_summary}
}
\end{center}
\end{table}

\subsection{Clifford orbits of maximal magic states}

As discussed in Section \ref{sec:magic}, Clifford unitaries preserve the SRE, ensuring that the SRE remains invariant across the orbit of states generated by these operations. Specifically, starting from a given fiducial state $| \psi_f \rangle$, the set of states produced by all possible Clifford unitaries—the Clifford orbit—maintains a constant SRE value, $M_2(| \psi_f \rangle)$.

\rev{For the one-, two-, and three-qubit maximal-magic sets considered here, the Clifford action is transitive \cite{Liu:2025frx,ohta_2025_15651358}.  The $E_6$-selected one-qutrit set behaves differently \cite{Appleby2005,TabiaAppleby2013}.}

The single-qutrit phase-modded Clifford group, ${\cal C}(1,3)/U(1)$, is generated by the Hadamard gate $H$ and the phase gate $S$, given by:
\begin{equation}
H=\frac{1}{\sqrt{3}}\left(\begin{array}{ccc}
1 & 1 & 1 \\
1 & \omega & \omega^2 \\
1 & \omega^2 & \omega
\end{array}\right), 
~~~S=\left(\begin{array}{ccc}
1 & 0 & 0 \\
0 & 1 & 0 \\
0 & 0 & \omega
\end{array}\right),
~~~\omega = e^{2 \pi i/3}.
\end{equation}

To investigate the Clifford orbits of maximal magic states, we applied all 216 elements of ${\cal C}(1,3)/U(1)$ to 45 maximal magic states derived from the second-shortest vectors of the $E_6$ lattice.

\begin{revision}
The reason for non-transitivity is already visible from the two stabilisers.  The 9-ray orbit is the Hesse SIC and has the enhanced projective stabiliser $\operatorname{SL}(2,3)$ of order 24; the other representative has stabiliser $C_6$ of order 6 and therefore an orbit of 36 rays.  The exceptional symmetry of the Hesse SIC agrees with Zhu's classification of doubly transitive SICs \cite{Appleby2005,TabiaAppleby2013,Zhu2015SuperSIC}.  Here double transitivity concerns the nine projectors within the Hesse SIC, whereas the non-transitivity at issue concerns the full set of 45 $E_6$-selected fiducials.  Points in one group orbit must have conjugate stabilisers and hence stabilisers of equal order, so the orders 24 and 6 prevent the two families from lying in a single orbit \cite{ohta_2025_15651358}.
\end{revision}

\subsection{On the correspondence between the $E_6$ shortest vectors and the 1-qutrit stabiliser states}
\label{sec:E6stab}
Thus far, we have shown that in certain quantum systems, the minimal and maximal magic states appear systematically via a direct mapping from the first- and second-shortest vectors of specific symmetric lattices. These results are derived from direct computations of the stabiliser R{\' e}nyi entropy; however, the underlying mechanism behind these correspondences remains unexplained.

One significant challenge is that the relevant lattice vectors reside in high-dimensional spaces, which complicates the analysis. Additionally, the sheer number of these lattice vectors further adds to the difficulty. 
However, when considering the correspondence between the $E_6$ first-shortest vectors and the single-qutrit stabiliser states, these challenges are less severe. 
After all, the single-qutrit system involves only 12 stabiliser states.

In this subsection, we explicitly construct the 12 qutrit stabiliser states and verify whether they are embedded in the $E_6$ lattice with a square norm of $\langle \vec{x}, \vec{x} \rangle = 3$.

In the single-qutrit system, all 27 elements of the Weyl-Heisenberg group, $\overline {\cal W}(1,3)$, are obtained by
\be
\overline {\cal W}(1,3) = \left\{ \tau^s D_{a_1, a_2} \mid s, a_1, a_2 \in \{0,1,2\} \right\}
\ee
with
\be
D_{a_1,a_2} \equiv \tau^{a_1 a_2} X^{a_1} Z^{a_2}
\ee
and $\tau = -e^{\frac{\pi i}{3}} = \omega^2$.
By taking the computational basis as $|1 \ket \cong (1,0,0)$, $|2 \ket \cong (0,1,0)$ and $|3 \ket \cong (0,0,1)$, operators
$D_{1,0} = X$, $D_{0,1} = Z$, $D_{1,1} = \tau X Z$ and $D_{1,2} = \tau^2 X Z^2$ are represented by the following matrices:
\begin{equation}
D_{1,0} \cong \begin{pmatrix}
0 & 1 & 0 \\
0 & 0 & 1 \\
1 & 0 & 0
\end{pmatrix}, \quad
D_{0,1} \cong \begin{pmatrix}
1 & 0 & 0 \\
0 & \omega & 0 \\
0 & 0 & \omega^2
\end{pmatrix}, \quad
D_{1,1} \cong \begin{pmatrix}
0 & 1 & 0 \\
0 & 0 & \omega \\
\omega^2 & 0 & 0
\end{pmatrix}, \quad
D_{1,2} \cong \begin{pmatrix}
0 & 1 & 0 \\
0 & 0 & \omega^2 \\
\omega & 0 & 0
\end{pmatrix}\,.
\label{WH-generators}
\end{equation}

The stabiliser groups are defined as the maximal abelian subgroups of $\overline {\cal W}(1,3)$ that do not contain any phase element other than 1.
Twelve such groups can be found, which are generated from a single generator as 
\begin{equation}
\begin{aligned}
\mathcal{S}_1  &= \left\langle D_{1,0} \right\rangle, &
\mathcal{S}_2  &= \left\langle \omega D_{1,0} \right\rangle, &
\mathcal{S}_3  &= \left\langle \omega^2 D_{1,0} \right\rangle, 
\\
\mathcal{S}_4  &= \left\langle D_{0,1} \right\rangle, &
\mathcal{S}_5  &= \left\langle \omega D_{0,1} \right\rangle, &
\mathcal{S}_6  &= \left\langle \omega^2 D_{0,1} \right\rangle, 
\\
\mathcal{S}_7  &= \left\langle D_{1,1} \right\rangle, &
\mathcal{S}_8  &= \left\langle \omega D_{1,1} \right\rangle, &
\mathcal{S}_9  &= \left\langle \omega^2 D_{1,1} \right\rangle,
\\
\mathcal{S}_{10} &= \left\langle D_{1,2} \right\rangle, &
\mathcal{S}_{11} &= \left\langle \omega D_{1,2} \right\rangle, &
\mathcal{S}_{12} &= \left\langle \omega^2 D_{1,2} \right\rangle.
\end{aligned}
\end{equation}
Let $\hat{s}_i$ denote the generator of the stabiliser group ${\cal S}_i$. Each stabilizer group consists of three elements, ${\cal S}_i = \{ 1, \hat{s}_i, \hat{s}_i^2 \}$.
Each ${\cal S}_i$ stabilises a unique state of single-qutrit, meaning that the stabiliser state is uniquely characterised by the condition:
\be
S_i | \Psi_i \ket = | \Psi_i \ket,~~~~\forall  S_i \in \{ 1, \hat s_i, \hat s^2_i \}\,.
\ee
To determine $| \Psi_i \rangle$, it suffices to identify the eigenvector of $\hat{s}_i$ corresponding to the eigenvalue $+1$.
From explicit calculations, we find the following 12 stabiliser states:  
\begin{equation}
\begin{aligned}
\left| \Psi_1 \right\rangle  &\cong \vec{S}_1  = (1,\ 1,\ 1), &
\left| \Psi_2 \right\rangle  &\cong \vec{S}_2  = (1,\ \omega,\ \omega^2), &
\left| \Psi_3 \right\rangle  &\cong \vec{S}_3  = (1,\ \omega^2,\ \omega), \\[1ex]
\left| \Psi_4 \right\rangle  &\cong \vec{S}_4  = (\theta,\ 0,\ 0), &
\left| \Psi_5 \right\rangle  &\cong \vec{S}_5  = (0,\ \theta,\ 0), &
\left| \Psi_6 \right\rangle  &\cong \vec{S}_6  = (0,\ 0,\ \theta), \\[1ex]
\left| \Psi_7 \right\rangle  &\cong \vec{S}_7  = (1,\ 1,\ \omega^2), &
\left| \Psi_8 \right\rangle  &\cong \vec{S}_8  = (1,\ \omega,\ \omega), &
\left| \Psi_9 \right\rangle  &\cong \vec{S}_9  = (1,\ \omega^2,\ 1), \\[1ex]
\left| \Psi_{10} \right\rangle &\cong \vec{S}_{10} = (1,\ 1,\ \omega), &
\left| \Psi_{11} \right\rangle &\cong \vec{S}_{11} = (1,\ \omega,\ 1), &
\left| \Psi_{12} \right\rangle &\cong \vec{S}_{12} = (1,\ \omega^2,\ \omega^2).
\end{aligned}
\label{stabilizer_list}
\end{equation}
Here, we have fixed the norm of the vectors to be $\bra \vec S_i, \vec S_i \ket = 3$, i.e.\ the length of the $E_6$ first-shortest vectors, and chose appropriate phases. 

The key question is whether these vectors indeed represent the coordinates of the $E_6$ first-shortest vectors. 
This is equivalent to determining whether
\be
\vec S_i = \vec c_i \cdot M_{E_6}, 
\label{E6_condition}
\ee
admits a solution within the domain $\vec c_i = {\mathbb Z}[\omega]^3$.

To address this, we take a direct approach. 
We find the following explicit solutions:\footnote{Note that $\omega^2 = -1 -\omega \in {\mathbb Z}[\omega]$.}
\begin{equation}
\begin{aligned}
\vec{c}_1   &= (0,\ 0,\ 1), &
\vec{c}_2   &= (-\omega,\ 1,\ \omega^2), &
\vec{c}_3   &= (\omega^2,\ -1,\ \omega), \\[1ex]
\vec{c}_4   &= (1,\ 0,\ 0), &
\vec{c}_5   &= (0,\ 1,\ 0), &
\vec{c}_6   &= (-1,\ -1,\ \theta), \\[1ex]
\vec{c}_7   &= (-\omega,\ -\omega,\ \omega^2), &
\vec{c}_8   &= (\omega^2,\ 0,\ \omega), &
\vec{c}_9   &= (0,\ \omega,\ 1), \\[1ex]
\vec{c}_{10} &= (\omega^2,\ \omega^2,\ \omega), &
\vec{c}_{11} &= (0,\ -\omega^2,\ 1), &
\vec{c}_{12} &= (-\omega,\ 0,\ \omega^2)\,.
\end{aligned}
\label{c_list}
\end{equation}

It is straightforward to verify that multiplying Eq.\ \eqref{E6_condition} by the phase factors $\{ \pm 1, \pm \omega, \pm \omega^2 \}$ yields six solutions corresponding to six phase-equivalent stabiliser states. 
Consequently, each stabiliser state generates six distinct $E_6$ lattice vectors with the square norm 3.
In total, this produces all $12 \times 6 = 72$ $E_6$ first-shortest vectors.

\section{Conclusion}
\label{concl}

In this work, we introduced a novel geometric framework that links symmetric lattices to the structure of magic states in low-dimensional quantum systems. 
By constructing a map from points in ${\mathbb R}^{2D}$ (or ${\mathbb C}^D$) to pure states in a $D$-dimensional Hilbert space, we systematically generated and analysed stabiliser and maximal magic states across several systems. 
Specifically, we recovered the complete sets of known stabiliser states for the two-qubit, three-qubit and single-qutrit systems, as well as the known set of maximal magic states for the two-qubit system. 
Furthermore, we uncovered new sets of maximal magic states for the three-qubit and single-qutrit systems, derived from the second-shortest vectors of the $BW_{16}$ and $E_6$ lattices, respectively. 
\rev{Based on the structure and symmetry of the $BW_{16}$-derived states, we conjecture that the 15,360 rays form the complete set of three-qubit maximal magic states; the qutrit count of 45 refers only to the second shell of $E_6$ \cite{TabiaAppleby2013,ohta_2025_15651358}.}

Our main results demonstrate that symmetric lattices can serve as powerful tools for cataloguing and generating quantum states with extremal magic properties. 
For the two-qubit system, we showed that all 240 first-shortest vectors of the $E_8$ lattice correspond to the full set of 60 stabiliser states, and that the 2,160 second-shortest vectors yield all 480 known maximal magic states. 
In the three-qubit system, we mapped the 4,320 shortest vectors of the $BW_{16}$ lattice to the complete set of 1,080 stabiliser states, and from the 61,440 second-shortest vectors, we identified 15,360 distinct maximal magic states. 
These results led us to conjecture that 15,360 is the total number of maximal magic states in the three-qubit system. 
\rev{For the single-qutrit system, we applied the same method to the $E_6$ lattice and recovered all 12 stabiliser states together with 45 lattice-selected maximal magic states, partitioned into two distinct Clifford orbits \cite{Appleby2005,TabiaAppleby2013,ohta_2025_15651358}.}

Beyond enumeration, we conducted a detailed analysis of the entanglement properties of the extremal states. In the three-qubit case, we classified both stabiliser and maximal magic states into entanglement classes based on one-to-one, one-to-other and genuine tripartite concurrence. This classification revealed distinct patterns: stabiliser states occupied three clearly defined entanglement classes, while maximal magic states formed two classes distinguished by the presence or absence of bipartite one-to-one entanglement. 

The implications of our findings are multifold. 
First, they offer a new avenue for constructing extremal magic states in a closed form, bypassing the need for brute-force numerical optimisation or group-theoretic enumeration. Second, the appearance of highly symmetric lattices such as $E_8$ and $BW_{16}$ in the structure of quantum magic provides compelling evidence of deeper algebraic relationships between quantum resource theory and lattice theory—connections that may be further elucidated through the lens of representation theory or algebraic number theory. 
Third, the ability to map extremal states using geometric considerations may provide practical advantages for quantum information protocols, such as magic state distillation, quantum circuit synthesis, and fault-tolerant state preparation.

Looking forward, several promising directions arise from this study. One avenue is to explore whether similar lattice-based constructions can be extended to higher-qubit systems or to hybrid systems involving qudits of varying dimensions. Another important direction is the investigation of suboptimal or intermediate magic states, whose geometric origins may reside in longer lattice vectors or in alternative lattice types. Additionally, the link between WH-covariant structures (such as SICs and MUBs) and the identified maximal magic states suggests that further exploration of group-covariant POVMs may yield new families of useful quantum resources.

In conclusion, this work bridges the domains of lattice theory and quantum information, offering both conceptual unification and computational practicality. The identification of extremal magic states via symmetric lattices not only enriches our theoretical understanding but also lays a foundation for practical algorithms in quantum computing and quantum state engineering. By continuing to probe this geometric perspective, we may uncover more of the hidden structure underlying quantum advantage.

\appendix

\section{Weyl-Heisenberg group and Stabiliser states for single-qudit}\label{SRE}

The stabiliser R{\' e}nyi entropies for the $n$-qubit system $({\mathbb C}^2)^{\otimes n}$ can be generalised to the general $n$-qudit system $({\mathbb C}^d)^{\otimes n}$ \cite{Wang2023,Cuffaro:2024wet}. 
In this subsection, we consider the single-qudit case with general $d$ and $n=1$.

Let ${\mathbb Z}_d = \{0, \cdots, d-1\}$ be the set of congruence classes modulo $d$.
Consider the orthonormal computational basis of the $d$-dimensional system $\{ | k \ket \}$ ($k \in {\mathbb Z}_d$). 
Define the shift $X$ and clock $Z$ operators as
\be
X | k \ket = | k+1 \ket,~~~~Z | k \ket = \omega^k | k \ket 
\ee
with $\omega = e^{\tfrac{2 \pi i}{d}}$ and $| d \ket \equiv | 0 \ket$.
They satisfy 
\be
X^d = Z^d = 1,~~~~
X^a Z^b = \omega^{- ab} Z^b X^a \,.
\ee
For $d=2$, $X$ and $Z$ are reduced to the Pauli matrices $\sigma_x$ and $\sigma_z$, respectively. 

From the shift and clock operators, the displacement operators are defined as \cite{PhysRevLett.116.040501}
\begin{equation}
D_{a_1,a_2} \equiv \tau^{a_1 a_2} X^{a_1} Z^{a_2}
\end{equation}
with $a_1, a_2 \in \mathbb{Z}_d$ and $\tau = -e^{\frac{\pi i}{d}}$. 
The displacement operators satisfy $\left(D_{a_1,a_2}\right)^d = 1$, the multiplication law
\be
D_{a_1,a_2}\, D_{b_1,b_2}
= \tau^{a_2 b_1 - a_1 b_2} D_{(a_1+a_2),(b_1+b_2)}
= \omega^{a_2 b_1 - a_1 b_2} D_{b_1,b_2}\, D_{a_1,a_2}
\,,
\label{commutation_relation}
\ee
the unitarity 
\be
D_{a_1,a_2}^\dagger = D_{-a_1,-a_2} = D_{a_1,a_2}^{-1}\,,
\ee
and the orthogonality relation 
\be
{\rm Tr} \left[ D_{a_1,a_2} D_{b_1,b_2}^\dagger \right] = d \, \delta_{a_1,b_1} \delta_{a_2,b_2}\,.
\ee

The single-qudit Weyl-Heisenberg group $\overline {\cal W}(1,d)$---the generalisation of the Pauli group in ${\mathbb C}^d$---is generated by the set of displacement operators as 
\be
\overline {\cal W}(1,d)
= \bra \{ D_{a_1,a_2} \} \ket = \{ \tau^s D_{a_1,a_2}   \}\,
\ee
with $a_1,a_2 \in {\mathbb Z}_d$ and
$s \in {\mathbb Z}_{d}$ $({\mathbb Z}_{2d})$ for odd (even) $d$.\footnote{Note that $\tau^d = 1$ when $d$ is odd, while $\tau^d = -1$ when $d$ is even.}

The Clifford group ${\cal C}(1,d)$ is defined as the normaliser of the WH group in the unitary group $U(d)$
\be
{\cal C}(1,d) = \{ U \in U(d) \,|\, U \overline {\cal O} U^\dagger \in \overline {\cal W}(1,d), ^\forall \overline {\cal O} \in \overline 
 {\cal W}(1,d)  \} .
\ee
The stabiliser states are defined as the states that are reachable from $|0 \ket$ by applying the Clifford unitaries. 

To see another aspect of the stabiliser states, define the set of stabiliser groups $\{ {\cal S}_i \}$ as the set of maximal abelian subgroups of $\overline {\cal W}(1,d)$ that do not contain any phase element other than 1.\footnote{If a non-trivial phase element is included in ${\cal S}_i$, Eq.\ \eqref{stabilizer_def2} will not admit any slution other than $| \Psi_i \ket = 0$.
In some literature, the stabiliser group is defined as the maximal abelian subgroup of the WH group that does not contain the $-1$ element. 
While the latter condition suffices to eliminate all non-trivial phase elements when $d$ is even, for odd $d$, it is necessary to explicitly require that ${\cal S}_i$ contains no non-trivial phase elements.}
Each stabiliser group $S_i$ uniquely defines the corresponding stabiliser state $|\Psi \ket_i$ by
\be
S_i | \Psi_i \ket = | \Psi_i \ket,~~~^\forall S_i \in {\cal S}_i\,.
\label{stabilizer_def2}
\ee

Analogously to the Pauli string, we define the quotient group ${\cal W}(1,d) = \overline {\cal W}(1,d)/\bra \tau^s \ket$, ignoring the phase. 
In other words, ${\cal W}(1,d) = \{ D_{a_1,a_2} \}$ with $a_1,a_2 \in {\mathbb Z}_d$. 
The stabiliser R{\' e}nyi entropy (SRE) for the single-qudit system is defined by Eq.\ \eqref{defM0} with
\be
\Xi_\alpha ( | \psi \ket )
\equiv \frac{1}{d} \sum_{ {\cal O} \in {\cal W}(1,d)}
\left| \bra \psi | {\cal O} | \psi \ket  \right|^{2 \alpha}\,,
\label{eq:Xi_app}
\ee
where the summation is taken over all elements of the phase-quotiented WH group, ${\cal W}(1,d)$.

A key property of the SRE is that it vanishes if and only if $| \psi \rangle$ is a stabiliser state, $| \psi \ket \in \{ | \Psi_i \ket \}$. 
Below, we calculate $\Xi_\alpha ( | \Psi_i \ket )$ for the stabiliser state associated with ${\cal S}_i$ and show that $\Xi_\alpha ( | \Psi_i \ket ) = 1$, implying $M_\alpha ( | \Psi_i \ket ) = 0$.
As a first step, we rearrange the summation in Eq.\ \eqref{eq:Xi_app} to initially sum over the $d$ elements of $\mathcal{W}(1,d) = \{ D_{a_1,a_2} \}$ that are included in $\mathcal{S}_i$, up to phase factors $\tau^s$.
We then evaluate the remaining sum over the $d^2 - d$ elements not contained in $\mathcal{S}_i$.
%
%
Namely, we have
\be
\Xi_\alpha ( | \Psi_i \ket )
= \frac{1}{d} \sum_{ S_i \in {{\cal S}_i}}
\left| \bra \Psi_i | S_i | \Psi_i \ket  \right|^{2 \alpha}
+
\frac{1}{d} \sum_{ {\cal O} \in \mathcal{W}(1, d) \setminus \mathcal{S}_i }
\left| \bra \Psi_i | {\cal O} | \Psi_i \ket  \right|^{2 \alpha}
\,,
\label{Xi_two_sums}
\ee
In the first term, we have removed the phase factors, as $|\tau^s|^{2\alpha} = 1$. 
Using Eq.\ \eqref{stabilizer_def2} and the fact that $|{\cal S}_i| = d$, the first term evaluates to 1.

We now show that the contribution of the second term is zero. 
Let $D_{b_1,b_2}$ be an element of ${\cal W}(1,d)$ that is not included in ${\cal S}_i$ modulo phases. 
Since ${\cal S}_i$ is a maximal abelian group, one can always find an element $S_i^{(s,a_1,a_2)} =\tau^s D_{a_1,a_2}$ in ${\cal S}_i$ that does not commute with $D_{b_1,b_2}$ (otherwise, $D_{b_1,b_2}$ must be included in ${\cal S}_i$).   
Using Eq.\ \eqref{commutation_relation}, this implies
\be
S_i^{(s,a_1,a_2)} \cdot D_{b_1,b_2} | \Psi_i \ket = \omega^{a_2 b_1 - a_1 b_2} D_{b_1,b_2} \cdot S_i^{(s,a_1,a_2)} | \Psi_i \ket = \omega^{a_2 b_1 - a_1 b_2} D_{b_1,b_2} | \Psi_i \ket
\ee
with $\omega^{a_2 b_1 - a_1 b_2} \neq 1$.
This implies that $D_{b_1,b_2} | \Psi_i \rangle$ is an eigenstate of the unitary operator $S_i^{(s,a_1,a_2)}$ with an eigenvalue different from 1.
Since eigenstates of a unitary operator corresponding to different eigenvalues are orthogonal, it follows that
$\langle \Psi_i | D_{b_1,b_2} | \Psi_i \rangle = 0$.
This argument holds for every element in the second summation, implying that the second term in Eq.\ \eqref{Xi_two_sums} vanishes.
This confirms that stabiliser states satisfy $\Xi_\alpha ( | \Psi_i \rangle ) = 1$, and thus their SRE vanishes, $M_\alpha ( | \Psi_i \rangle ) = 0$.

\section{Generator matrices}
\label{app:generators}

We provide the generator matrices of the $E_8$ and $BW_{16}$ lattices used in this paper.

\begin{equation}
M_{E_8} \,=\,
\left(\begin{array}{rrrrrrrr}
1 & -1 & 0 & 0 & 0 & 0 & 0 & 0 \\
0 & 1 & -1 & 0 & 0 & 0 & 0 & 0 \\
0 & 0 & 1 & -1 & 0 & 0 & 0 & 0 \\
0 & 0 & 0 & 1 & -1 & 0 & 0 & 0 \\
0 & 0 & 0 & 0 & 1 & -1 & 0 & 0 \\
0 & 0 & 0 & 0 & 0 & 1 & -1 & 0 \\
-\frac{1}{2} & -\frac{1}{2} & -\frac{1}{2} & -\frac{1}{2} & -\frac{1}{2} & -\frac{1}{2} & \frac{1}{2} & \frac{1}{2} \\
0 & 0 & 0 & 0 & 0 & 1 & 1 & 0
\end{array}\right)\,,
\label{E_8_gen}
\end{equation}

\begin{equation}
M_{BW_{16}} \,=\, \frac{1}{2}\left(\begin{array}{llllllllllllllll}
1 & 1 & 1 & 1 & 1 & 1 & 1 & 1 & 1 & 1 & 1 & 1 & 1 & 1 & 1 & 1 \\
0 & 2 & 0 & 0 & 0 & 0 & 0 & 2 & 0 & 0 & 0 & 2 & 0 & 2 & 0 & 0 \\
0 & 0 & 2 & 0 & 0 & 0 & 0 & 2 & 0 & 0 & 0 & 2 & 0 & 0 & 2 & 0 \\
0 & 0 & 0 & 2 & 0 & 0 & 0 & 2 & 0 & 0 & 0 & 2 & 0 & 0 & 0 & 2 \\
0 & 0 & 0 & 0 & 2 & 0 & 0 & 2 & 0 & 0 & 0 & 0 & 0 & 2 & 2 & 0 \\
0 & 0 & 0 & 0 & 0 & 2 & 0 & 2 & 0 & 0 & 0 & 0 & 0 & 2 & 0 & 2 \\
0 & 0 & 0 & 0 & 0 & 0 & 2 & 2 & 0 & 0 & 0 & 0 & 0 & 0 & 2 & 2 \\
0 & 0 & 0 & 0 & 0 & 0 & 0 & 4 & 0 & 0 & 0 & 0 & 0 & 0 & 0 & 0 \\
0 & 0 & 0 & 0 & 0 & 0 & 0 & 0 & 2 & 0 & 0 & 2 & 0 & 2 & 2 & 0 \\
0 & 0 & 0 & 0 & 0 & 0 & 0 & 0 & 0 & 2 & 0 & 2 & 0 & 2 & 0 & 2 \\
0 & 0 & 0 & 0 & 0 & 0 & 0 & 0 & 0 & 0 & 2 & 2 & 0 & 0 & 2 & 2 \\
0 & 0 & 0 & 0 & 0 & 0 & 0 & 0 & 0 & 0 & 0 & 4 & 0 & 0 & 0 & 0 \\
0 & 0 & 0 & 0 & 0 & 0 & 0 & 0 & 0 & 0 & 0 & 0 & 2 & 2 & 2 & 2 \\
0 & 0 & 0 & 0 & 0 & 0 & 0 & 0 & 0 & 0 & 0 & 0 & 0 & 4 & 0 & 0 \\
0 & 0 & 0 & 0 & 0 & 0 & 0 & 0 & 0 & 0 & 0 & 0 & 0 & 0 & 4 & 0 \\
0 & 0 & 0 & 0 & 0 & 0 & 0 & 0 & 0 & 0 & 0 & 0 & 0 & 0 & 0 & 4
\end{array}\right)\,,
\label{BW_star_gen}
\end{equation}

\section{Search range}
\label{app:search}

To determine the complete set of first- and second-shortest vectors of a given lattice $\Lambda = \textsf{Span}_{\mathbb{Z}} \, M$ in $d$-dimensions, one must solve the quadratic form equation
$\vec{a} \cdot G \cdot \vec{a}^{\,T} = \ell_n$,
for $\vec{a} \in \mathbb{Z}^d$, where $M$ is the generator matrix of the lattice, $G \equiv M M^T$ is the associated Gram matrix, and $\ell_n$ denotes the known squared norm of the $n$-th shortest lattice vectors.
When performing a numerical search for solutions, it is useful first to identify the range of values that the components $a_i \in {\mathbb Z}$ may take.
In this section, we derive a formula for that range. Specifically, we show that each component $a_i$ must satisfy the bound
\begin{equation}
|a_i| \leq \left\lfloor \sqrt{ \ell_n (G^{-1})_{ii} } \right\rfloor,
\label{range_formula}
\end{equation}
where $(G^{-1})_{ii}$ denotes the $i$-th diagonal element of the inverse of the $G$ matrix, and $\lfloor \cdot \rfloor$ denotes the floor function.

To determine the maximal and minimal values that a particular component $a_i$ can attain, subject to the norm constraint, we employ the method of Lagrange multipliers. Define the objective function
$f_i(a_1, \ldots, a_d) = a_i$,
and impose the constraint
\be
g(a_1, \ldots, a_d; \ell_n) = a_k G_{kl} a_l - \ell_n = 0,
\ee
where summation over repeated indices is implicit. The corresponding Lagrange function is defined by
\be
\mathcal{L}_i(a_1, \ldots, a_d, \lambda) \,=\, f_i(a_1, \ldots, a_d) + \lambda \, g(a_1, \ldots, a_d; \ell_n)\,.
\ee

The stationary points are found by solving
\be
\frac{\partial \mathcal{L}}{\partial a_1} = 0, \quad \cdots, \quad \frac{\partial \mathcal{L}}{\partial a_d} = 0, \quad \frac{\partial \mathcal{L}}{\partial \lambda} = 0.
\ee
The first $d$ conditions yield the linear system
\be
\frac{\partial \mathcal{L}}{\partial a_k} = \delta_{ik} + 2\lambda \, G_{kl} a_l = 0,
\ee
which leads to the solution
\begin{equation}
a_k = -\frac{(G^{-1})_{ki}}{2\lambda}.
\label{eq:ak}
\end{equation}
Substituting them into the last equation $g(a_1,\ldots,a_d;\ell_n) = 0$, we find
\be
\lambda^2 = \frac{(G^{-1})_{ii}}{4\ell_n}.
\ee
Finally, eliminating $\lambda$ in Eq.\eqref{eq:ak} for $k = i$, we obtain the extremal value of $a_i$, and since $a_i \in \mathbb{Z}$, we arrive at the bound given in Eq.~\eqref{range_formula}.

In practical applications, this formula offers an efficient method for collecting all numerical solutions. 
Below we list the diagonal elements of $G^{-1}$ used in this paper for the $E_8$ and $BW_{16}$ lattices:
\bea
(G_{E_8}^{-1})_{ii} &=&
\left( 2,6,12,20,30,14,4,8\right),
\nonumber \\
(G_{BW_{16}}^{-1})_{ii} &=& 
\left( 4,2,2,2,2,2,2,8,2,2,2,8,2,8,8,8 \right)\,.
\eea

To apply this method to the $E_6$ lattice defined in $\mathbb{C}^3$, we embed the lattice into $\mathbb{R}^6$ by treating the imaginary part of each complex coordinate as an independent orthogonal direction. 
In this way, each component of the $\mathbb{C}^3$ vector contributes two real dimensions, allowing the lattice to be fully represented in $\mathbb{R}^6$. Within our construction, the coordinates of the $E_6$ lattice are given by:
\be
\vec c \,=\, (\beta_1, \b_2, \b_3) 
\begin{pmatrix}
\theta & 0 & 0 \\
0 & \theta & 0 \\
1 & 1 & 1 
\end{pmatrix},~~~
\theta = i \sqrt{3},
\ee
with $\beta_i = a_i + \omega b_i$, $\omega = e^{2 \pi i/3}$ and $a_i,b_i \in {\mathbb Z}$.
Using the inverse of the map $\varphi: {\mathbb R}^6 \to {\mathbb C}^3$ defined in Eq.\ \eqref{map_phi}, we represent $\vec x$ in ${\mathbb R}^6$ as 
\be
\vec c \xlongrightarrow{\varphi^{-1}} 
\vec x = 
( {\rm Re}[\theta \beta_1 + \beta_3], {\rm Re}[\theta \beta_2 + \beta_3], {\rm Re}[\beta_3],
{\rm Im}[\theta \beta_1 + \beta_3], {\rm Im}[\theta \beta_2 + \beta_3], {\rm Im}[\beta_3]
),
\ee
This vector can also be written as
\be
\vec x = (a_1,a_2,a_3,b_1,b_2,b_3) \cdot M_{E_6}^{\rm real}
\ee
with
\begin{equation}
M^{\rm real}_{E_6} = \frac{1}{2}\left(\begin{array}{rrrrrr}
0 & 0 & 0 & 2 \sqrt{3} & 0 & 0 \\
0 & 0 & 0 & 0 & 2 \sqrt{3} & 0 \\
2 & 2 & 2 & 0 & 0 & 0 \\
-3 & 0 & 0 & -\sqrt{3} & 0 & 0 \\
0 & -3 & 0 & 0 & -\sqrt{3} & 0 \\
-1 & -1 & -1 & \sqrt{3} & \sqrt{3} & \sqrt{3}
\end{array}\right)\,.
\label{real_E_6}
\end{equation}
With this expression, the search ranges for $a_i, b_i \in {\mathbb Z}$ can be found by using the formula \eqref{range_formula} with
\begin{equation}
\left[ \left( G_{E_6}^{\rm real} \right)^{-1} \right]_{i i}
=\frac{1}{9}(8,8,12,8,8,12)\,.
\end{equation}

\section{Entanglement measures}
\label{app:entangle}

\subsection{Bipartite entanglement measures}

In this work, we use {\it concurrence} to quantify entanglement. 
For a bipartite pure state $|\psi_{AB} \rangle \in {\cal H}_A \otimes {\cal H}_B$, the concurrence is defined as
\be
\mathcal{C}(|\psi_{AB}\rangle) = \sqrt{2 \left(1 - \operatorname{Tr}(\varrho_A^2)\right)},
\label{conc_pure}
\ee
where $\varrho_A = \operatorname{Tr}_B(\rho_{AB})$ is the reduced density matrix of subsystem $A$. The definition is symmetric under the exchange of subsystems $A$ and $B$. Concurrence reaches its maximum value $\mathcal{C} = 1$ for maximally entangled two-qubit states.

For mixed states $\rho$, concurrence is extended via the convex roof construction \cite{Uhlmann1998}:
$\mathcal{C}(\rho) = \inf_{\{p_k, |\psi_k\rangle\}} \sum_k p_k\, \mathcal{C}(|\psi_k\rangle)$,
where the infimum is taken over all decompositions $\rho = \sum_k p_k |\psi_k\rangle\langle\psi_k|$.

An analytical formula for $\mathcal{C}(\rho)$ exists in the two-qubit case \cite{Wootters1998}:
\be
\mathcal{C}(\rho) = \max(0, \eta_1 - \eta_2 - \eta_3 - \eta_4),
\label{conc_mix}
\ee
where $\eta_i$ are the eigenvalues (in decreasing order) of
$\sqrt{ \sqrt{\rho}\, (\sigma_y \otimes \sigma_y)\, \rho^*\, (\sigma_y \otimes \sigma_y)\, \sqrt{\rho} }$,
with $\rho^*$ denoting complex conjugation in the computational basis.

\subsection{Tripartite entanglement measures}

For a three-qubit pure state $| \Psi \ket \equiv | \psi_{ABC} \ket \in {\mathbb C}^2 \otimes {\mathbb C}^2 \otimes {\mathbb C}^2$, entanglement can be categorised into three types:

\begin{description}

\item[1. One-to-one entanglement:]

This refers to the entanglement between two individual qubits, such as between $A$ and $B$.
It is computed by tracing out the third qubit and applying the concurrence formula for mixed states \eqref{conc_mix}:
\be
{\cal C}_{AB} = {\cal C}(\varrho_{AB}), \quad \varrho_{AB} = \operatorname{Tr}_C(|\Psi \rangle\langle \Psi|).
\label{CAB}
\ee
The measures $\mathcal{C}_{AC}$ and $\mathcal{C}_{BC}$ are obtained similarly.
These pairwise concurrences can be nonzero even for bi-separable states.
For example, $\mathcal{C}_{AB} \ne 0$ for states of the form $| \gamma \rangle_C \otimes | \delta \rangle_{AB}$, but vanish for $| \alpha \rangle_A \otimes | \delta \rangle_{BC}$ and $| \beta \rangle_B \otimes | \delta \rangle_{AC}$.

\item[2. One-to-other entanglement:]

This captures the entanglement between one qubit and the remaining two, such as between $A$ and $BC$.
We treat the full pure state $|\Psi\rangle \in \mathcal{H}_A \otimes \mathcal{H}_{BC}$ as a bipartite system and apply the pure-state concurrence formula \eqref{conc_pure}:
\[
\mathcal{C}_{A(BC)} = \sqrt{2\left(1 - \operatorname{Tr}(\varrho_{BC}^2)\right)}, \quad
\varrho_{BC} = \operatorname{Tr}_A(|\Psi\rangle\langle\Psi|).
\label{Ci(jk)}
\]
The measures $\mathcal{C}_{B(AC)}$ and $\mathcal{C}_{C(AB)}$
are defined analogously.
One-to-other concurrence may also be nonzero for bi-separable states.
For example, $\mathcal{C}_{A(BC)} \ne 0$ for $| \beta \rangle_B \otimes | \delta \rangle_{AC}$ and $| \gamma \rangle_C \otimes | \delta \rangle_{AB}$, but vanishes for $| \alpha \rangle_A \otimes | \delta \rangle_{BC}$.

\item[3. Genuine tripartite entanglement:]

This form of entanglement vanishes for all fully separable and bi-separable states, making it a reliable indicator of genuine tripartite entanglement (GTE). A particularly well-behaved measure of GTE was recently introduced in Ref.\cite{Jin2023}, defined in terms of the geometric properties of a “concurrence triangle”, where each side of the triangle corresponds to a one-to-other concurrence (see Fig.\ref{fig:F3}).
The area of this triangle serves as a quantitative measure of GTE:
\be
\mathcal{F}_3 = \frac{4}{\sqrt{3}} \, \sqrt{
Q \left(Q - \mathcal{C}_{A(BC)}\right)
\left(Q - \mathcal{C}_{B(AC)}\right)
\left(Q - \mathcal{C}_{C(AB)}\right)},
\label{F3}
\ee
where $Q = \frac{1}{2} \left[ \mathcal{C}_{A(BC)} + \mathcal{C}_{B(AC)} + \mathcal{C}_{C(AB)} \right]$.
The normalization factor ensures $\mathcal{F}_3 \in [0,1]$.

\begin{figure}[t!]
 \centering
 \includegraphics[keepaspectratio, scale=0.08]
      {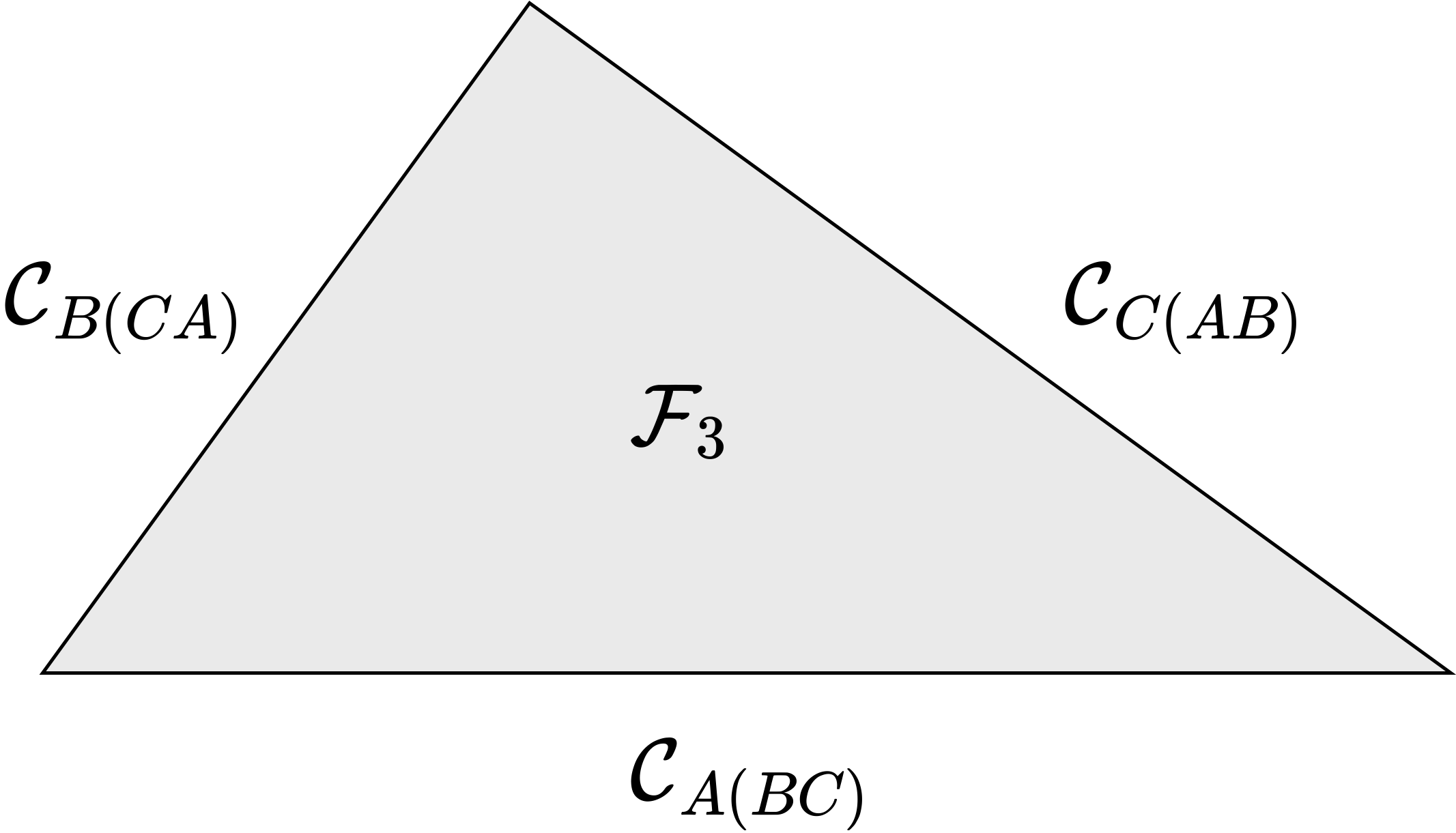}
 \caption{\small The genuinely tripartite entanglement measure $\mathcal{F}_3$ as the area of the concurrence triangle.
}
 \label{fig:F3}
\end{figure}

\end{description}


\bibliographystyle{JHEP}
\bibliography{refs} 

\end{document}